\documentclass[fleqn,usenatbib]{mnras}
\usepackage{newtxtext,newtxmath}
\usepackage[T1]{fontenc}
\usepackage{ae,aecompl}

\usepackage{graphicx}	
\usepackage{amsmath}	
\usepackage{amssymb}	

\title[deepCool: Fast and Accurate Cooling Rates]{deepCool: Fast and Accurate Estimation of Cooling Rates in Irradiated Gas with Artificial Neural Networks}

\author[T. P. Galligan et al.]{Thomas P. Galligan,$^{1}$\thanks{E-mail: thomas.galligan@bnc.ox.ac.uk}, Harley Katz,$^{1}$, Taysun Kimm$^2$, Joakim Rosdahl$^3$, 
\newauthor Jeremy Blaizot$^3$, Julien Devriendt$^1$, \& Adrianne Slyz$^1$
\\
$^1$Astrophysics, University of Oxford, Denys Wilkinson Building, Keble Road, Oxford OX1 3RH, UK\\
$^2$Department of Astronomy, Yonsei University, 50 Yonsei-ro, Seodaemun-gu, Seoul 03722, Republic of Korea \\
$^3$Univ Lyon, Univ Lyon1, Ens de Lyon, CNRS, Centre de Recherche Astrophysique de Lyon UMR5574, F-69230, Saint-Genis-Laval, France \\
}

\date{Accepted XXX. Received YYY; in original form ZZZ}

\pubyear{2018}

\begin{document}
\label{firstpage}
\pagerange{\pageref{firstpage}--\pageref{lastpage}}
\maketitle

\begin{abstract}
Accurate models of radiative cooling are a fundamental ingredient of modern cosmological simulations.  Without cooling, accreted baryons will not efficiently dissipate their energy and collapse to the centres of haloes to form stars.  It is well established that local variations in the amplitude and shape of the spectral energy distribution of the radiation field can drastically alter the cooling rate.  Here we introduce {\small deepCool}, {\small deepHeat}, and {\small deepMetal}: methods for accurately modelling the total cooling rates, total heating rates, and metal-line only cooling rates of irradiated gas using artificial neural networks.  We train our algorithm on a high-resolution cosmological radiation hydrodynamics simulation and demonstrate that we can predict the cooling rate, as measured with the photoionisation code {\small CLOUDY}, under the influence of a local radiation field, to an accuracy of $\sim5\%$.  Our method is computationally and memory efficient, making it suitable for deployment in state-of-the-art radiation hydrodynamics simulations.  We show that the circumgalactic medium and diffuse gas surrounding the central regions of a galaxy are most affected by the interplay of radiation and gas, and that standard cooling functions that ignore the local radiation field can incorrectly predict the cooling rate by more than an order of magnitude, indicating that the baryon cycle in galaxies is affected by the influence of a local radiation field on the cooling rate.
\end{abstract}




\section{Introduction}
In the context of galaxy formation, radiative cooling is one of the fundamental processes that allows gas to collapse to the centres of dark matter haloes and form stars.  Currently, most state-of-the art cosmological simulations implement radiative cooling, especially for metal-lines, by using pre-computed look-up tables that can be interpolated on-the-fly based on local properties of the simulation \citep[e.g.][]{Dubois2014,Volgelsberger2013,Crain2015}.  Early attempts at modelling the cooling function relied on a set of simplifying assumptions, such as that the gas is in collisional ionisation equilibrium and that the abundance ratios of individual elements are consistent with solar values \citep[e.g.][]{Cox1969,Raymond1976,Cox1971,Dalgarno1972,Shull1982,Sutherland1993}.  The latter approximation is certainly untrue as it is well established that different enrichment processes (e.g. different types of supernova explosions and stellar winds) yield abundances with ratios that are distinctly non-solar \citep[e.g.][]{Woosley1995,Nomoto1997a,Nomoto1997b,Umeda2002}.  Some modern cosmological simulations have begun to model the enrichment of individual elements \citep[e.g.][]{Kobayashi2006,Wiersma2009b,Volgelsberger2013,Dubois2014,Crain2015}, and crucially, many employ cooling tables computed on an element-by-element basis \citep{Wiersma2009,Gnat2012}.

The assumption that gas in the Universe can be modelled as being in collisional ionisation equilibrium (CIE) is also generally a poor one.  For low-density gas in the intergalactic medium (IGM), in the post-reionization epoch, the majority of the volume is permeated by UV background radiation \citep{Haardt1996,Haardt2012}, while at high densities in the interstellar medium (ISM), the proximity effect to local sources of radiation may also make CIE a poor approximation.  \cite{Efstathiou1992} demonstrated that for gas of primordial composition, the radiative cooling rate can be strongly suppressed by the presence of UV radiation (see their Figure 1).  Similarly, \cite{Wiersma2009} showed that the UV background can also affect metal-line cooling, suppressing the net cooling rates by up to an order of magnitude.  X-rays from accreting AGN or binaries are also expected to regulate the cooling rates \citep{Cantalupo2010,Kannan2016}.  Most state-of-the-art cosmological simulations now employ cooling tables that are computed under the presence of a cosmological UV background \citep[e.g.][]{Katz1996,Shen2010,grackle} although some simulations have begun to move beyond this and include the effects of local star formation \citep[e.g.][]{Kannan2014}.  While the assumption of a spectrally uniform UV background may be adequate for the IGM, different classes of sources (e.g. stars vs. AGN, different metallicity stars, different mass stars, etc.) have different luminosities and spectral energy distributions (SEDs) so local parcels of gas in the ISM may be subject to vastly different radiation fields.  \cite{Gnedin2012} explicitly demonstrated that the cooling function is very sensitive to the shape of the SED.  

Ideally, one would want to drop the approximation that the gas is in equilibrium and compute vast reaction networks for primordial species, metals, and molecules and directly solve the cooling rate on-the-fly in the simulation.  \cite{Richings2014a,Richings2014b} provide one of the most complete models in the literature for this in the context of cosmological simulations and indeed find non-trivial effects on the cooling function based on their non-equilibrium approach.  Complete and accurate models of cooling and heating processes require additional information such as dust content \citep{Bakes1994,Weingartner2001}, cosmic rays \citep{Wentzel1971}, and magnetic fields \citep{Dolag2008}. 

Even with the addition of a non-trivial SED to the computation of the metal-line only cooling function, the memory requirement of the cooling tables needed to model this effect becomes nearly computationally prohibitive \citep{Gnedin2012} as modern supercomputers only contain $\sim2-8$~Gb of RAM per core.  If one wanted to pre-compute the effects of a varying radiation field on an element-by-element basis for all relevant species, at multiple temperatures and densities, with enough resolution to accurately model the cooling function, this would certainly exceed the memory limit of most MPI systems.  For this reason, the high dimensional cooling tables are often compressed into fitting functions which are extremely cheap to store and fast to execute \citep[e.g.][]{Rosen1995,Wolfire2003}.  \cite{Gnedin2012} were the first to create cooling tables in the context of a sufficiently arbitrary radiation field and demonstrated that the modification to the cooling rate can be encapsulated by three free parameters that are based on normalised photoionisation rates of HI, HeI, CVI, and in the Lyman-Werner band.  This work is particularly timely as cosmological simulations that include on-the-fly radiation transfer have become significantly more common \citep[e.g.][]{Gnedin2014,Oshea2015,Pawlik2017,Katz2018a,Katz2018b}.  However, \cite{Gnedin2012} emphasise that the cooling functions can only be ``approximated'' with their model as specific cases suffer from ``catastrophic'' errors, which are almost certainly unavoidable when trying to compress information from a very high dimensional space.  

In this work we develop a technique for accurately modelling the cooling function of astrophysical gas subject to an arbitrary radiation field.  Since the problem consists of accurately compressing high-dimensional cooling tables into a sufficiently low-memory and fast executable, we employ artificial neural networks \citep[ANNs,][]{ANNs}. 
The approach we present in this paper will be demonstrated on a cosmological radiative transfer simulation to show that ANNs can accurately encapsulate the cooling functions of gas irradiated by a varying SED.  Furthermore, this technique is not limited to only radiation fields as it can easily be extended to take into account non-solar abundance ratios, cosmic rays, and any other physics that may affect the cooling rate.  Likewise, all simulations using this method are completely reproducible. 

This paper is organised as follows.  In Section~\ref{Methods}, we introduce the machine learning algorithms that we train to model the radiation-dependent cooling rates.  In Section~\ref{results}, we show how a trained ANN can accurately model the total cooling function, heating function, and metal-line-only cooling function for cells in a high-resolution cosmological radiation hydrodynamics simulation.  We quantify the errors on the method as a function of gas properties.  Finally, in Sections~\ref{discussion} \& \ref{conclusions}, we present a discussion of how we intend the algorithm to be deployed, quantify computational cost and accuracy in an isolated disk case, and present alternative use cases as well as our conclusions.

\section{Numerical Methods}
\label{Methods}
In order to teach our algorithm to estimate the cooling rates of gas at various temperatures, densities, metallicities and radiation fields, we create a ``training'' set by post-processing an output from a high-resolution cosmological radiation hydrodynamics zoom-in simulation of a massive galaxy during the epoch of reionization with the photoionisation code {\small CLOUDY}\footnote{Note that {\small CLOUDY} is our preferred choice of spectral synthesis code but in practice any other sophisticated thermochemistry code that can take into account a varying radiation field will suffice for our project.  The important point is that, in general, these codes tend to be prohibitively slow to embed into a cosmological simulation while ANNs are considerably faster, with only a small decrease in accuracy.} \citep{Ferland2017} to derive the total cooling rates, total heating rates, and metal-line only cooling rates under the influence of a local, inhomogeneous radiation field.  While the outputs we have generated with {\small CLOUDY} are subject to certain simplifying assumptions (e.g. we neclect dust, molecules, cosmic rays, and magnetic fields), since we aim to reproduce {\small CLOUDY} results as best as possible in our simulations, we refer to them as the ground truth. By training our algorithm on simulation data, we are able to employ it for all future simulations of a similar type.

\subsection{Simulation Details}\label{sim_details}

The exact details of the simulation we use for demonstration in this work are described in \cite{Katz2018}. In short, we have performed a cosmological radiation hydrodynamics zoom-in simulation of a massive Lyman-break galaxy that has mass $M_{\rm vir}\sim10^{11.8}{\rm M_{\odot}}$ at $z=6$.  We have used the publicly available adaptive mesh refinement (AMR) code {\small RAMSES-RT} \citep{Teyssier2002,Rosdahl2013,Rosdahl2015}.  Initial conditions are generated with {\small MUSIC} \citep{Hahn2011} in a box with a side length of 50~comoving~Mpc, using the following cosmology: $h=0.6731$, $\Omega_{\rm m}=0.315$, $\Omega_{\rm b}=0.049$, $\Omega_{\Lambda}=0.685$, $\sigma_8=0.829$, and $n_s=0.9655$, consistent with \cite{Planck2016}.  A set of high-resolution particles and gas cells are generated in the initial conditions around the Lagrange region of the massive halo giving an effective resolution of $4096^3$ particles in the zoom-in region, corresponding to a DM particle mass of $4\times10^4\ {\rm M_{\odot}}h^{-1}$.

We have used the same version of {\small RAMSES} as presented in \cite{Katz2017} and \cite{Kimm2017} to track non-equilibrium chemistry of H$_2$, H, H$^+$, He, He$^+$, He$^{++}$, and e$^-$, which are coupled to the radiation in eight frequency bins: infrared, optical, Habing, Lyman-Werner, HI-ionising, H$_2$ ionising, HeI-ionising, and HeII-ionising, via photoionisation, photo-heating, and UV and IR radiation pressure.  The radiation is advected across the grid using a moment method with the M1 closure for the Eddington tensor \citep{Levermore1984}.  Gas cooling is modelled for primordial species in a non-equilibrium fashion (see \citealt{Rosdahl2013}) while cooling from metal-lines in this simulation are currently tabulated at $T>10^4$ K using results from {\small CLOUDY} \citep{Ferland1998} using the CIE approximation and rely on analytical fits at lower temperatures \citep{Rosen1995}. 

The cells in the simulation adaptively refine when they contain either a dark matter mass $>3.2\times10^5 \,{\rm M_{\odot}}h^{-1}$ or a baryonic mass $>6\times10^4\,{\rm M_{\odot}}h^{-1}$, or when the local Jeans length is unresolved by at least four cells.  The simulation refines to a maximum level of 19, maintaining an approximate constant physical resolution of 13.6 pc, enough to begin resolving the escape channels for ionising radiation out of the interstellar medium \citep{Rosdahl2018} and into the IGM.

For our work here, we focus on the $\sim 850,000$ cells in a 5~kpc radius surrounding the most massive galaxy in the $z=10$ snapshot of the simulation.  Crucially for our experiment, the inhomogeneous radiation field that is expected to permeate the galaxy is modelled so that we can take into account this key feature to measure the radiation-affected cooling rates in post-processing.  The only source of radiation in the simulation comes from star particles.  We use a Schmidt law \citep{Schmidt1959} to model star formation which associates the star formation rate to the local free-fall time, the gas density, and an efficiency parameter.  Only cells with $\rho_{\rm gas}>100\,{\rm cm^{-3}}$ that have an unresolved thermo-turbulent Jeans length (see \citealt{Kimm2017}) are allowed to form stars.  If a cell is flagged for star formation, the number of particles formed is drawn from a Poisson distribution with a minimum stellar mass of 1000\,M$_{\odot}$ using the efficiency based on the local dynamical gas properties \citep{Kimm2017,Rosdahl2018}.  For the first 50~Myr of the lifetime of a star particle, it can explode via supernovae (SN), each event releasing $10^{51}$ ergs.  The occurrences of these events are randomly drawn from a delay-time distribution.  We use the mechanical feedback model for momentum injection presented in \cite{Kimm2015,Kimm2017,Rosdahl2018} and have boosted the number of SNe per M$_{\odot}$ by a factor of four compared to a standard Kroupa stellar IMF to overcome over-cooling \citep{Rosdahl2018}.  20\% of the mass fraction of each star particle is recycled back into the gas assuming a metallicity of the ejecta of 0.075 \citep{Kroupa2001}.  

Once a star is formed, it ejects radiation into its host cell on every fine time step of the simulation.  The age, mass, and metallicity of the particles are used to determine the shape and normalisation of the SED based on the {\small BPASSv2} model \citep{Stanway2016,Eldridge2008}, assuming a maximum stellar mass of 300\,M$_{\odot}$.  In this way, the inhomogeneous radiation field in the galaxy is naturally established by the distribution of star-forming regions.     

\subsection{Measuring the Cooling Rates}
As described earlier, we extract the $\sim850,000$ cells in the 5 kpc (40\% of the virial radius) region surrounding the most massive galaxy at $z=10$ for post-processing.  In the post-reionization epoch, much of the intergalactic medium sees a homogeneous and evolving UV background and this is relatively straightforward to add to cooling tables and is already commonplace in modern cosmological simulations \citep[e.g.][]{Wiersma2009}.  Thus, most interesting for us is the region in the proximity zone around galaxies where the local, time-varying radiation field dominates as well as in the regions of the ISM where the radiation field is strong enough such that CIE is no longer a good approximation.

For each cell, we run a {\small CLOUDY} model using the temperature, density, metallicity, and radiation field from the simulation.  We assume the GASS abundance ratios from \cite{Grevesse2010} as implemented into {\small CLOUDY} as well as a constant temperature approximation.  Since the radiation energy bins in the simulation are sampled quite coarsely, we assume that within an energy bin, the shape of the SED is consistent with that of the Milky Way interstellar radiation field \citep[ISRF,][]{Black1987} but the total normalisation of the SED within an energy bin is consistent with the flux of photons seen by the cell in the simulation.  For our current models, we only use six of the eight radiation bins in the simulation that impact the cooling rates (Habing: 5.6 -- 11.2 eV, Lyman-Werner: 11.2 -- 13.6 eV, HI-ionizing: 13.6 -- 15.2 eV, H$_2$ ionising: 15.2 -- 24.59 eV, HeI-ionising: 24.59 -- 54.42 eV, and HeII-ionising: $>54.42$ eV) while neglecting the optical and infrared as there are much larger uncertainties on these bins in the simulation due to coarse sampling.  We use {\small CLOUDY} to set up a thin slab (i.e. stop zone 1) with an open geometry such that the flux and spectrum of the radiation impinging the cloud is consistent with the flux of radiation in the simulation cell.  Note that we do not include the effects of dust, molecules (e.g. H$_2$), magnetic fields, or cosmic rays in our calculations\footnote{In principle, these can all be taken into account in future work using the methods presented in this paper.}.  For each cell, we output the total cooling rate, total heating rate, and metal-line only cooling rate.

\begin{figure*}
\centerline{\includegraphics[scale=1,trim={0 0 0 0},clip]{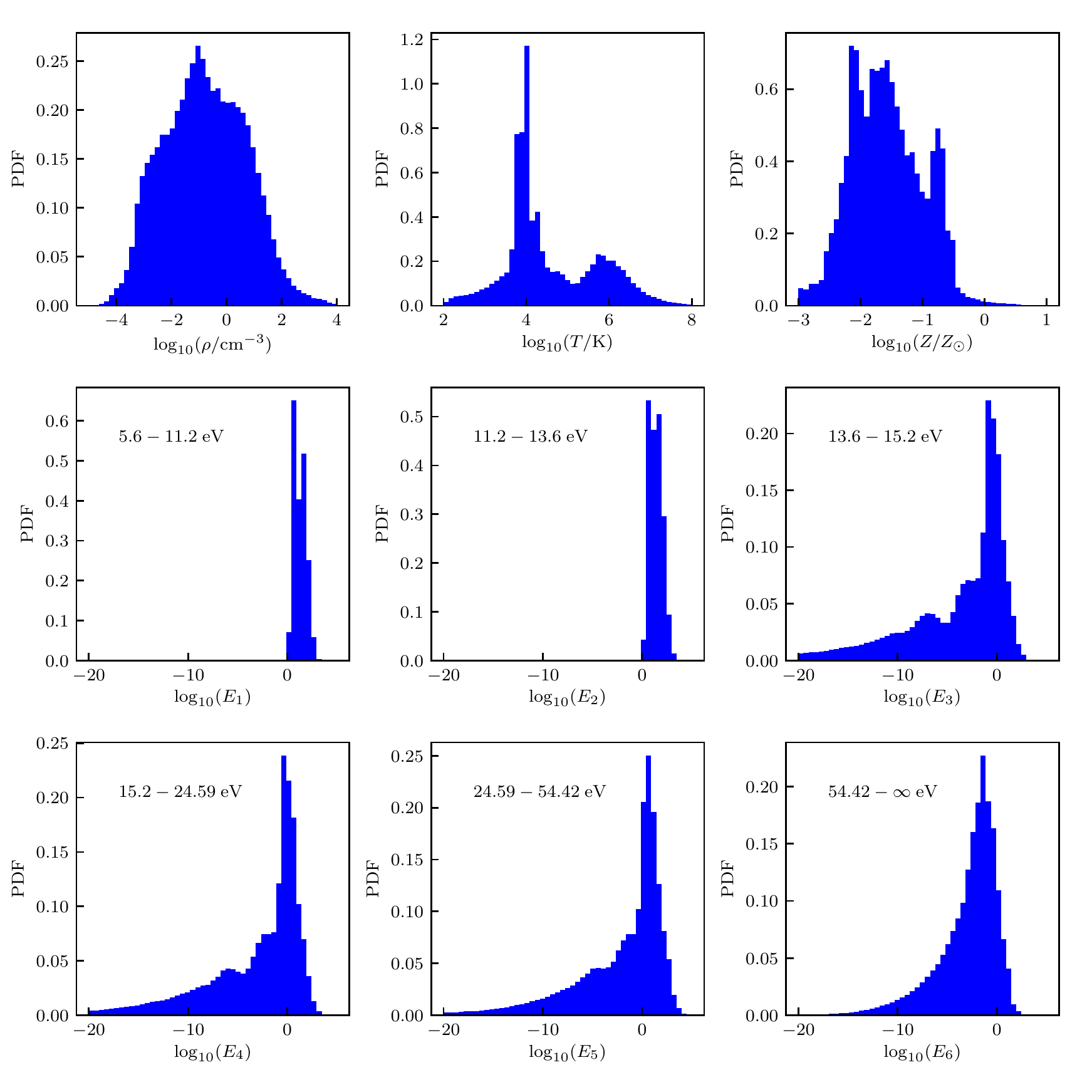}}
\caption{Histograms of the features used to train the machine learning algorithms. $E_1 - E_6$ correspond to the 6 energy bins used to model the radiation field and $\rho$ represents the hydrogen number density of the cell.  The energy ranges for each radiation bin are shown on each plot.  The value of $E_1 - E_6$ represent the ratio of the total flux in that energy bin in the simulation with respect to the Milky Way interstellar radiation field within the same energy range. The histograms are normalised to have unit integral and represent a probability distribution function (PDF).}
\label{featurehists}
\end{figure*}

In total there are nine free parameters in our cooling model: temperature, density, metallicity, and the scale factor of the Milky Way ISRF spectrum in each of the six radiation bins that we use.  In Figure~\ref{featurehists}, we show normalised histograms for each of these parameters for the $\sim 850,000$ cells post-processed with {\small CLOUDY}.  The gas densities are spread over more than eight orders of magnitude ($\sim10^{-4}-10^4{\rm \,cm^{-3}}$) with a peak at $\sim0.1{\rm \, cm^{-3}}$.  The temperatures range from $10^2$ -- $10^8$ K with a peak at $10^4$ K roughly corresponding to the ionisation temperature of hydrogen where the cooling rate significantly decreases towards lower temperatures \citep[e.g.][]{Wiersma2009}. As our current algorithm is trained on a simulation where most cells are at sub-solar metallicity, our current algorithm is more applicable to high-redshift systems.  The radiation intensities in the Habing and Lyman-Werner bands tend to be, on average, much higher than what is found in the Milky Way ISM.  Note that our simulation is relatively optically thin in these two bands and since the galaxy is at $z=10$, its size is much smaller than an equivalent mass galaxy at $z=0$ which drastically increases the intensity.  In the bands with energies higher than 13.6 eV, we tend to find a peak within an order of magnitude of the Milky Way interstellar radiation field.  Some cells scatter to much higher values and are indeed close to young star-forming regions.  There is an extreme tail in the distribution towards low radiation fluxes, stretching nearly 20 orders of magnitude, which covers the regime of completely self-shielded cells without any radiation.  These cells should already be well modelled with the CIE approximation.  Finally, we also point out that our simulations do not contain any strong source of X-ray photons such as AGNs or X-ray binaries and having this additional component may have a strong affect on the cooling rate as shown in \cite{Cantalupo2010,Gnedin2012,Kannan2016}.

\subsection{Cooling Rate Estimators}
We aim to develop an estimator for the total cooling, total heating, and metal-line cooling rates that is accurate enough for use in state-of-the-art cosmological simulations.  In addition to accuracy, the algorithm must satisfy computational constraints so that it is extremely fast to execute and requires limited space in memory to be competitive with other estimators, such as those tabulated \citep{Wiersma2009,grackle,Gnedin2012} or analytical functions \citep[e.g.][]{Rosen1995} which are not computationally demanding.  This is indeed difficult given the high dimensionality of the problem.  For this reason, we test four machine learning algorithms described below.

In order to train each algorithm, we have split our dataset into a ``training'' and ``testing'' group such that 80\% of the data ($\sim680,000$ cells) is used to optimise the algorithm while the other 20\% ($\sim170,000$ cells) is withheld to determine how well the algorithm generalises to unseen data.  The performance on the ``test'' set is crucial for determining how the algorithm will perform when deployed in a simulation.  Each algorithm also has its own set of hyper-parameters, such as architecture of the network, number of trees in the forest, or number of neighbours.  In order to optimise these, during training, we split the ``training'' set once more into a new ``training'' set and a ``validation'' set in a 90\%-10\% split.  We then train the algorithm with a given set of hyper-parameters on the new ``training'' set and test them on the ``validation'' set.  We vary the hyper-parameters and pick the model that performs best on the ``validation'' set as our final model for that algorithm.

We have rescaled the data so that the mean of each input parameter (temperature, density, etc.) is zero and the standard deviation is one.  The scaling factors are computed using only the ``training'' set and applied to the ``test'' when predictions were made so that no information about the ``test'' set was available during training.  We optimised each algorithm by minimising the mean squared error on the log of the cooling rates so that the algorithm was not biased to preferentially perform better on cells with higher cooling rates.  Since the distribution of some features is strongly peaked (for instance the strong peak at $10^4$ K, i.e. Lyman-alpha cooling), in principle, the algorithms may end up being biased towards getting the cooling more correct at these temperatures than others because a large proportion of the cells in the ``training'' set are at or around this temperature.  This is, for instance, potentially problematic for the lowest ($T\lesssim10^2$ K) and highest ($T\gtrsim10^8$ K) temperatures where there are very few cells to train on.  In principle, one could circumvent this bias by either up-sampling (i.e. weight more heavily) the cells at these under-represented regions of feature space or down-sampling the over-represented values so that the accuracy of the algorithm is less biased between different regions of feature space.  One could also sample based on the contribution of metal-line cooling to the total cooling rate, although this depends on metallicity as for very metal enriched gas, metal-line cooling could be dominant up to $T\sim10^7$ K (see Section~3.2).  In this work, however, we have opted not to take this approach as we much prefer to have the most accurate cooling rate predictions possible for the most cells in the simulation (which would require neither up- nor down-sampling) while also having a handle on where the algorithm may fail.  This is further discussed in Section~\ref{results}.  

Having described how the algorithms are trained and how the hyper-parameters are optimised we now describe the exact methods that we test.

\subsubsection{Artificial Neural Networks}
While often cited for usage as a classification tool, artificial neural networks (ANNs) are equally valid for regression problems \citep{ANNs}.  These algorithms can be extremely memory friendly (our algorithm only requires $\sim40$ kb of memory) and computationally efficient with the down side being that they can take a while to train.  They tend to generalise well if trained properly.  For these reasons, ANNs are potentially an ideal tool for use with cosmological simulations. 

\begin{table*}
\centering
\begin{tabular}{@{}lcccccc@{}}
\hline
Model & MSE$_{\Lambda}$ & MSE$_{\Lambda}$ & MSE$_{\Gamma}$ & MSE$_{\Gamma}$ & MSE$_{\Lambda_{\rm metal}}$ & MSE$_{\rm \Lambda_{\rm metal}}$ \\
 & (train) & (test) & (train) & (test) & (train) & (test) \\
\hline
Artificial Neural Network: {\small deepCool, deepHeat, deepMetal} & 0.00317 & 0.00331 & 0.00442 & 0.00465 & 0.00256 & 0.00273 \\
Random Forest Regression & 0.00024 & 0.00169 & 0.00024 & 0.00176 & 0.00022 & 0.00189\\
Nearest Neighbours Regression & 0.0 & 0.05190 & 0.0 & 0.01932 & 0.0 & 0.01374 \\
Optimised Nearest Neighbours Regression & 0.0 & 0.04305 & 0.0 & 0.01764 & 0.0 & 0.01370 \\
\hline
\end{tabular}
\caption{Mean squared errors (MSE) on the total cooling rates, total heating rates, and metal-line only cooling rates for each of the four algorithms tested for both the training and test sets. As expected, there is a small loss in accuracy between the training and test sets; however, this is relatively small for the ANN and Random Forest, indicating that these models generalise well.}
\label{ebins}
\end{table*}

We use a standard feed-forward neural network\footnote{The algorithm was designed and trained with {\small KERAS} (https://keras.io/) using a {\small THEANO} backend (http://deeplearning.net/software/theano/index.html).} and our architecture consists of an input layer of 9 dimensions (representing the number of input parameters: temperature, density, metallicity and the six radiation parameters), two hidden layers of 20 neurons each, and an output layer of one neuron\footnote{We have tested networks with between 15 and 30 neurons in each hidden layer.}.  The number of neurons in each hidden layer was optimised to obtain the lowest mean squared error on the ``validation'' set.  The weights and biases of each neuron in the hidden and output layers were initialised with a normal distribution with a mean of zero and a standard deviation of 0.05.  In each hidden layer, the weights and biases are passed through an activation function before proceeding to the next layer and for this we use a ReLU activation \citep{relu}.  The weights and biases on each neuron are optimised using backpropogation to calculate gradients of the mean squared error cost function.  The ADAM optimiser is employed to update these quantities \citep{adam}.  The algorithm is trained for $\sim1000$ epochs using a batch size of 10,000 cells ($\sim1.3\%$ of the ``training'' set).  The score on the ``validation'' set is computed on-the-fly during training, and the model with the best score on the ``validation'' set is saved.

\subsubsection{Random Forests}
Random forests are an example of an ensemble method that can also be used for both classification and regression \citep{Ho1995,Breiman2001}.  They use a combination of decision trees where each tree is trained on a sub-sample of the original dataset which is the same size as the original but sampled with replacement (i.e. bootstrapping).  The value predicted by the random forests is the average value of all of the decision trees.  Unlike ANNs, random forests can be extremely memory-intensive; our random forest requires $\sim 28-40$ Gb of memory. Despite this drawback, they offer many attractive features. One such feature is the ability to extract so-called ``feature-importances'' from the training data. These importances describe how heavily each feature is involved in the decision process which gives an indication of the physics driving the cooling rates.

We use the ensemble random forest regressor implementation in {\small scikit-learn}\footnote{http://scikit-learn.org/stable/} without setting a maximum depth of the trees so that they can become as arbitrarily deep as needed for the most accurate prediction.  We vary the number of trees in the forest from 100 to 800 and use cross-validation to settle on 500 as the optimal number of trees for the algorithm.

\begin{figure*}
\centerline{\includegraphics[scale=1,trim={0 0 0 0},clip]{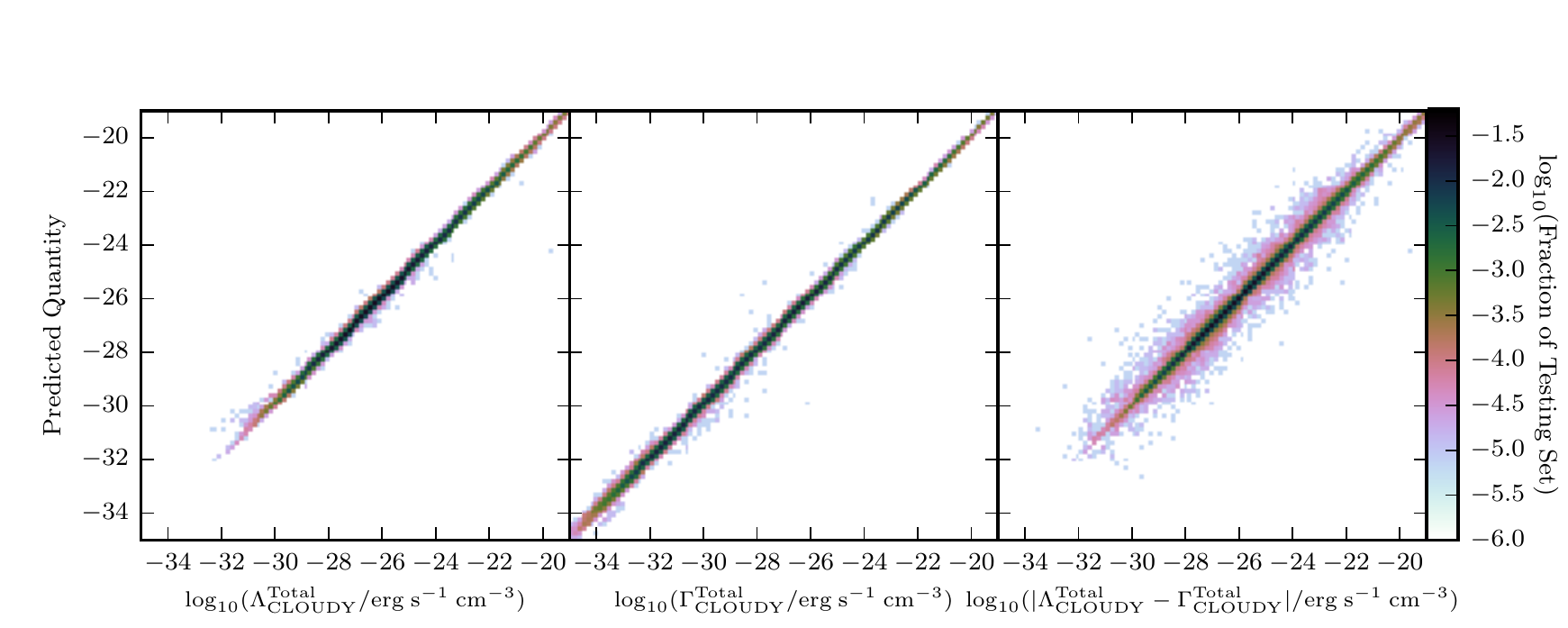}}
\caption{2D log histogram of the total {\small CLOUDY} cooling rate ($\Lambda_{\rm CLOUDY}^{\rm Total}$, left), total heating rate ($\Gamma_{\rm CLOUDY}^{\rm Total}$, centre), and the difference between the two (i.e. the total net cooling rate, right) verses the predicted values from {\small deepCool} and {\small deepHeat} for cells in the test set.  The colour of the cell represents the fraction of cells in the test set at that location.  In all cases, the artificial neural networks predict cooling and heating rates that are generally consistent with the {\small CLOUDY} values and hence fall very close to the 1-to-1 line.  There are no obvious systematics present in the predictions.}
\label{predictions2d}
\end{figure*}

\subsubsection{Nearest Neighbours}\label{knn}
Nearest Neighbours regression is a fairly straightforward algorithm that uses a kernel-weighted average of the values of N-nearest neighbours in the feature space to make predictions.  This algorithm is an example of a ``lazy learner'' as much of the computational expense occurs at run-time when a prediction is made.  For this reason it is generally inexpensive to train a nearest neighbours regressor and it depends on how quickly a tree can be built for an efficient neighbour search.  In addition to choosing the number of neighbours to average over and the kernel, one must define a distance metric.  Since we have scaled the data, using a Euclidean metric implies that all features have the same importance for determining the cooling rates.  This is clearly unlikely to be true.  Furthermore, as the dimensions increase, the distance metric may become less informative.  Because the neighbour search must be performed at run-time, the tree built for the neighbour search must be kept in memory at run-time making this algorithm less attractive for cosmological simulations due to the memory expense.  Nevertheless, given its simplicity, it is worth testing and we optimise the number of neighbours used for the regression in the range 1 to 10 and have used cross-validation to settle on 4.  We have weighted each of the N-neighbours by the inverse distance from the target point.

\subsubsection{Optimised Nearest Neighbours}
Optimised nearest neighbours is an adaptation of the previous algorithm that we devised where we use the feature importance calculation given by the random forest regressor to optimise the distance metric used in the nearest neighbours regressor. Given an $n$-dimensional feature space (in our case $n=9$), the random forest generates a set of feature importances $\{\alpha_1, \dots ,\alpha_n\}$, where $\alpha_i \in [0,1]$ and $\sum_i \alpha_i = 1$. Given these importances, we can modify the distance metric to
\begin{equation}
	d(\mathbfit{x},\mathbfit{y}) \equiv \sqrt{\sum_{i=1}^n (x_i-y_i)^2} \  \longrightarrow \  \sqrt{\sum_{i=1}^n \alpha_i (x_i-y_i)^2},
\end{equation}
where $\mathbfit{x}$ and $\mathbfit{y}$ are coordinate vectors in our feature space, with $\mathbfit{x}$ representing the new set of parameters with which we want to predict the cooling rate, and $\mathbfit{y}$ being a feature vector in our training set. 

This modification encourages the algorithm to preferentially select neighbouring points that have small differences in the most important features, and to care less if these points have a large difference in less important features. As a consequence, we expect this algorithm to perform slightly better than a standard nearest neighbours approach, because in principle, it has knowledge of the physics of the system.  We average the result over the 4 nearest neighbours using the inverse distance as the weighting scheme in the regression.

\section{Results}
\label{results}
Comparing the results of the four different algorithms will give an indication of which is the best to embed into a cosmological simulation.  In Table~\ref{ebins}, we present the mean squared errors on the training and test sets for all four algorithms for the total cooling rate (MSE$_{\Lambda}$), total heating rate (MSE$_{\Gamma}$), and metal-line only cooling rates (MSE$_{\Lambda_{\rm metal}}$).  Note that we have trained the algorithm on the logarithm of the rates so the mean squared errors are in log-scale.  Comparing the results between the training and test datasets provides intuition on how well the model is expected to perform when applied to unseen data.  

Overall, the random forest regressor tends to perform best for all three quantities on the test sets, outperforming the neural networks by a factor of two on the total cooling rates, a factor of 2.6 on the total heating rates, and a factor of 1.4 on the metal-line cooling rates.  In all cases the nearest-neighbour regressor performs at least an order of magnitude worse compared to the random forest regressor.  Note that on the training sets, the nearest-neighbour regressor will always have a mean squared error of zero because the nearest-neighbour will have a distance of 0 giving it an infinite weight.  The optimised nearest neighbours improves the performance slightly due to the change in metric, but not enough to make it competitive with either the ANN or the random forest.

The difference between the performance of the ANNs on the training set versus test set is extremely small, indicating that the algorithm we have trained generalises very well.  There is a much larger difference between the training and test scores for the random forest, due to the architecture of the algorithm.

Ideally, since the random forest has the lowest errors on all of the test sets, we would directly export this into a cosmological simulation.  However, the memory needed to store the 500 trees is extremely large, $\sim30$ Gb, which is much too memory intensive for even the most state-of-the-art distributed memory machines.  Furthermore, the random forest has a much longer execution time for making predictions compared to the ANNs which makes it less competitive for numerical simulations.  Thus we must make a compromise between accuracy and numerical constraints.  For this reason, we focus the remainder of our analysis on the performance of the ANNs; however, we suggest that if cooling and heating rates are required for post-processing simulations as we have done here, random forests may be the better algorithm to use given the increase in accuracy.  In either case, both algorithms are orders of magnitude faster than directly running {\small CLOUDY} for millions of cells (for example, the ANNs are roughly seven orders of magnitude faster).

\subsection{Total Cooling and Heating Rates}
In Figure~\ref{predictions2d}, we show a 2D histogram of the total cooling rates, total heating rates, and the absolute value of the difference between the two on the test set as calculated with {\small CLOUDY} versus what is predicted by {\small deepCool} and {\small deepHeat} using an ANN.  Remarkably, nearly all points fall on the 1-to-1 line indicating that the algorithm is performing extremely well, independent of the value of the cooling or heating rate.  The distribution of heating rates has a larger range than the cooling rates which makes them slightly more difficult to predict. This is reflected in Table~\ref{ebins}, where the mean squared error is 0.00442 on the test set for the heating rates, versus 0.00317 for the cooling rates.

Although we used the mean squared error in $\log_{10}(\Lambda^{\rm Total})$ and $\log_{10}(\Gamma^{\rm Total})$ to train {\small deepCool} and {\small deepHeat}, it is insightful to measure the absolute value of the fractional differences between our models and what is calculated with {\small CLOUDY}.  For {\small deepCool}, 50\% of all cells in the test set have an error $<6.27\%$ while this value is $6.56\%$ for {\small deepHeat}.  The 90th and 99th percentiles show an error of $19.8\%$ and $40.1\%$ for {\small deepCool} and $22.8\%$ and $50.6\%$ for {\small deepHeat}.  While in most of our cells, we can accurately predict the total cooling and heating rates to better than $\sim20\%$, there are a few cells that scatter to higher errors due to the limitations of trying to compress all of the information needed to accurately model the cooling rates into the ANN.  

\begin{figure}
\centerline{\includegraphics[scale=1,trim={0 0 0 0},clip]{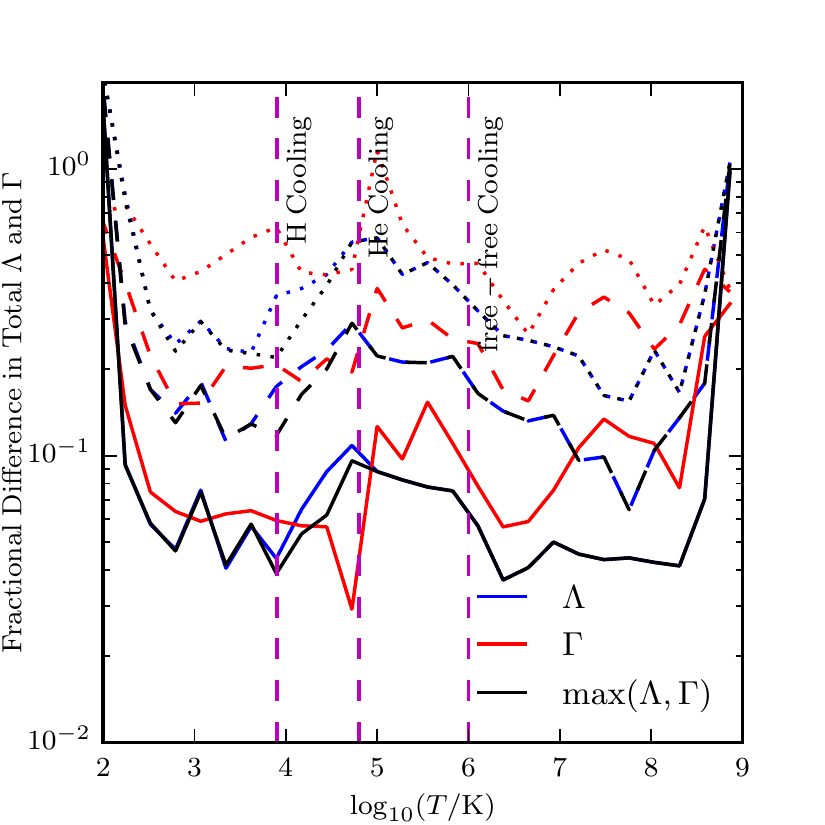}}
\caption{Fractional difference, $\frac{|\Lambda_{\rm deepCool,deepHeat}-\Lambda_{\rm CLOUDY}^{\rm Total}|}{\Lambda_{\rm CLOUDY}^{\rm Total}}$, between the predicted total cooling (blue) and heating (red) rates and those generated by {\rm CLOUDY} for the test set as a function of temperature.  The solid, dashed, and dotted lines represent the 50th, 90th, and 99th percentiles in the fractional difference at fixed temperature.  The black lines show the same quantities but for the maximum between the cooling and heating rates for a given cell.  The vertical magenta lines show the approximate temperatures where H, He, and Bremsstrahlung cooling dominate in a metal free gas.}
\label{frac_diff}
\end{figure}

In Figure~\ref{frac_diff}, we plot the fractional difference between the total cooling and heating rates predicted by {\small deepCool} and {\small deepHeat} with the ones generated by {\small CLOUDY} at the 50th, 90th, and 99th percentiles as a function of temperature.  For most of the range in temperature our model accurately predicts the cooling and heating rates.  There is a steep increase in error at the lowest and highest temperature ranges in this Figure.  This is due to the fact that we have very few cells at these temperatures in our simulations.  Note that low-temperature cells tend to be at very high densities and not in the IGM as our training set was selected from the inner regions of a massive galaxy.  There are already strong systematic uncertainties in this regime due to inaccuracies in subgrid models of dust, molecules, and feedback and thus the error in our cooling models is small compared to the uncertainty that these models introduce.  Furthermore, our star formation prescription is sensitive to the thermo-turbulent Jeans length \citep{Kimm2017} and in the low-temperature regime this quantity and the efficiency of star formation per unit free-fall time are more sensitive to the gas velocities than the temperature.  Hence small errors in the cooling rate are unlikely to affect the star formation history of the system.  Once a star particle forms, the gas will quickly heat up due to photoheating.  At very high temperatures, $T\sim10^9$ K, there is also a factor of two difference but Bremsstrahlung cooling is very efficient at these temperatures and this can actually be computed analytically as is common in simulations.

At $10^{2.5}<T/{\rm K}<10^{8.5}$ we find no strong biases in the predicted rates.  The model struggles slightly more in predicting both the heating and cooling rates at $T\sim10^5$ K; however, there is also a dip in number of cells at this temperature (see Figure~\ref{featurehists}).  This is because this is a thermally unstable region in temperature-space because of a peak in the cooling curve \citep[e.g.][]{Katz1996}.  Because there are fewer cells to train on, the 50th percentile fractional difference in the cooling and heating rates at this temperature is $\sim10\%$ rather than $\sim6\%$.  In any case, our results are very competitive with the accuracy of the \cite{Gnedin2012} model for the cooling functions of irradiated gas.  Our model has more free parameters than the one presented in \cite{Gnedin2012} which is the main reason we had to employ machine learning instead of a table.

\subsection{Metal-Line Cooling Rates}
It is often the case that in cosmological simulations, the cooling and heating rates for hydrogen, helium, their ions, and their associated molecules can be solved for relatively inexpensively without requiring that they are in equilibrium.  This is indeed the case in {\small RAMSES-RT} and directly calculating the non-equilibrium cooling in the simulation is more accurate than using tabulated values that are created with an external photoionisation code (e.g. {\small CLOUDY}).  Since solving for the cooling rates from primordial species separately from the metal-line cooling rates is a common technique in numerical simulations, it is prudent for us to create a model for fast estimation of the cooling rates due to metal-lines.

For each gas cell in our set, we have measured the cooling rates due to metals\footnote{The included metals are: Li, Be, B, C, N, O, F, Ne, Na, Mg, Al, Si, P, S, Cl, Ar, K, Ca, Sc, Ti, V, Cr, Mn, Fe, Co, Ni, Cu, Zn} and trained an additional neural network, {\small deepMetal}, using the same test-train-validation split as was previously used for the total heating and cooling rates.  Over the entire test set, we find that the median fractional difference between the predicted metal-line cooling rates and the actual metal-line cooling rates measured with {\rm CLOUDY} is 3.8\%.  This value increases to 10.5\% and 29.4\% in the 90th and 99th percentiles respectively.  In other words, we can predict the metal-line cooling rates of 99\% of the cells in the simulation to better than 30\% accuracy.

\begin{figure}
\centerline{\includegraphics[scale=1,trim={0 0 0 0},clip]{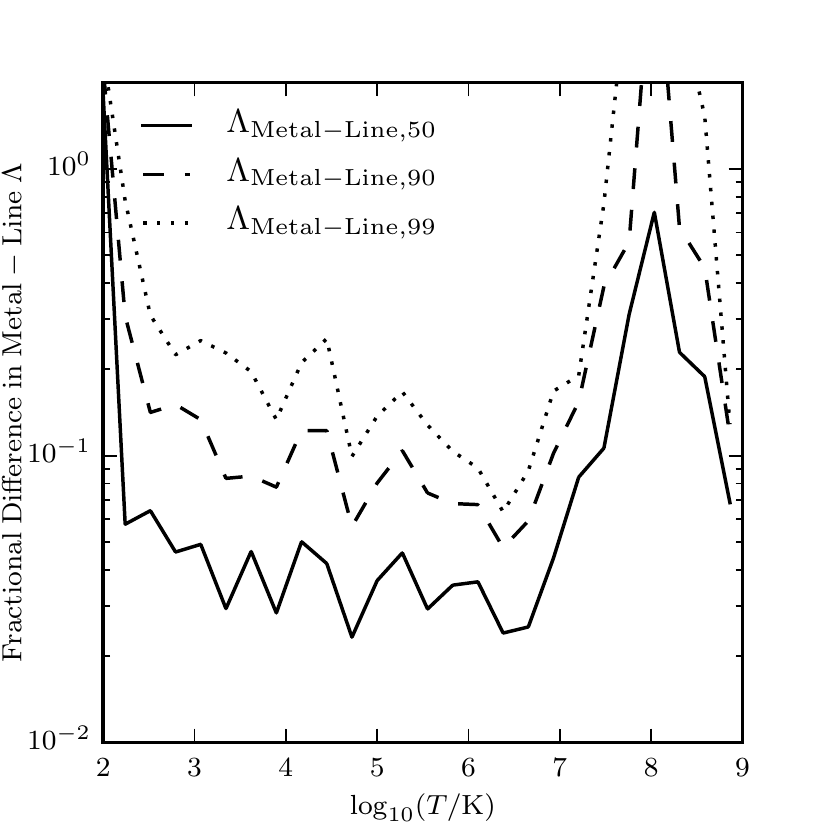}}
\caption{Fractional difference  $\frac{|\Lambda_{\rm deepMetal}-\Lambda_{\rm CLOUDY}^{\rm Metal}|}{\Lambda_{\rm CLOUDY}^{\rm Metal}}$ between the predicted metal-line cooling rates and the true cooling rates generated by {\rm CLOUDY} for the test set as a function of temperature.  The solid, dashed, and dotted lines represent the 50th, 90th, and 99th percentiles in the fractional difference respectively. }
\label{frac_diff_metal}
\end{figure}

In Figure~\ref{frac_diff_metal}, we plot the fractional difference between the predicted and ``true'' ({\small CLOUDY}) metal-line cooling rates as a function of gas temperature for the 50th, 90th, and 99th percentile of the distributions.  Our most accurate predictions exist in the temperature range $3<\log_{10}(T/{\rm K})<7$, where in general, the accuracy of our predictions is well within 20\% of the true cooling rate.  At very low temperatures, $T\sim 10^2$ K, we once again find that the error in our prediction rapidly rises which is due to the fact that we have very few cells in our simulation in this regime to train the algorithm on.  

\begin{figure}
\centerline{\includegraphics[scale=1,trim={0 0 0 0},clip]{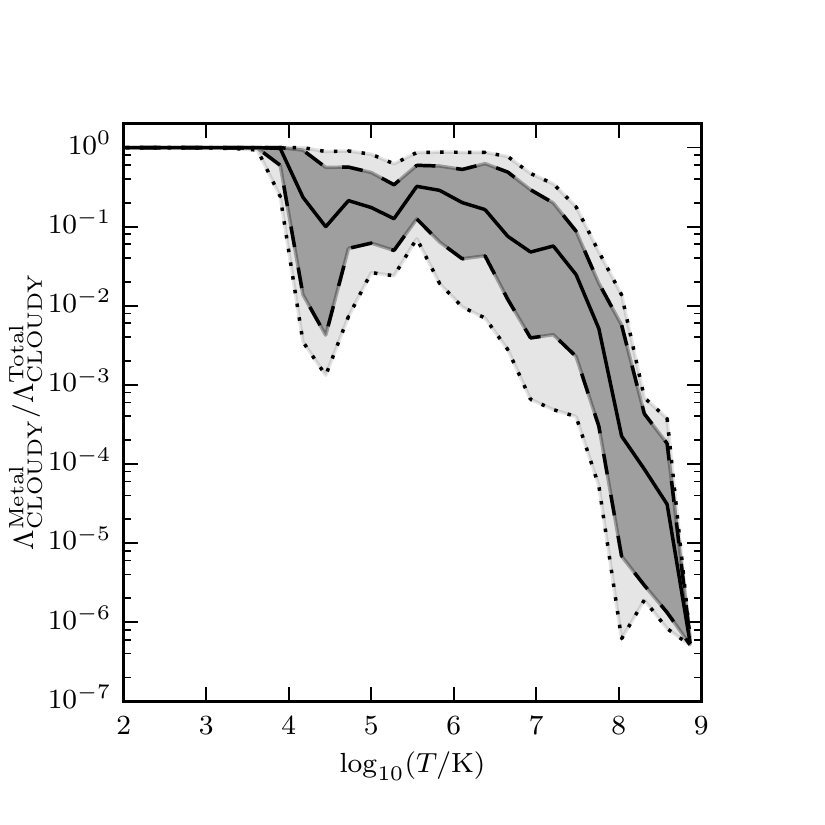}}
\caption{Ratio of metal-line cooling rate to total cooling rate as a function of temperature for all cells in the set as computed by {\small CLOUDY}.  The solid black line shows the median ratio while the dark grey and light grey bands represent cells within the 10th-90th and 1st-99th percentiles respectively.  The dispersion in the relation is due to differences in density, metallicity, and radiation field at a fixed temperature.  At temperatures $T\lesssim10^4$~K, metal-line cooling dominates the total cooling rate.  The contribution from metal-lines decreases with increasing temperature to effectively negligible values for $T\gtrsim10^{7.5}$~K.  Note that we have run {\small CLOUDY} without molecules so in reality, H$_2$ will also contribute to the low temperature cooling and the current cooling is all from atomic species (e.g. [CII]).  For low metallicity gas, this may be the dominant contribution at $100\ {\rm K}<T<10^4$~K.}
\label{mtrat}
\end{figure}

At high temperatures, $T\sim 10^8$ K, we also find ``catastrophic'' errors in our ability to predict the metal-line cooling rates as the 50th percentile difference between the predicted and true metal-line cooling rates at this temperature is $\sim70\%$.  At first glance, this may seem worrying given the large error in the prediction.  However, in Figure~\ref{mtrat}, we plot the ratio of metal-line cooling rate to the total cooling rate as a function of temperature for all cells in the set.  The solid black line shows the median relation while the light grey shaded region represents the bounds on this ratio between the 1st and 99th percentiles.  At temperatures, $T\sim 10^8$ K, the median contribution of metal-lines to the total cooling rate is 0.01\% suggesting that they are effectively negligible in this temperature regime.  This is because the total cooling rate at these temperatures is dominated by Bremsstrahlung.  Even if the error on our metal-line cooling prediction at this temperature was a factor of 10, this would only translate to a $\sim0.1\%$ error in the total cooling rate.  Hence the 70\% median uncertainty our model provides at this temperature is not a worrying feature of our model.  

At temperatures $T\lesssim10^7$ K, the contribution from metal-line cooling rates can reach 10\% of the total cooling rate for $\sim1\%$ of cells in the simulation (see Figure~\ref{mtrat}).  Referring back to Figure~\ref{frac_diff_metal}, by this temperature, the median error in the metal-line cooling rates is $\sim5\%$ indicating that in the regime where metal-line cooling actually impacts the temperature state of the gas, our {\small deepMetal} algorithm performs as expected.   

\begin{figure}
\centerline{\includegraphics[scale=1,trim={0 0 0 0},clip]{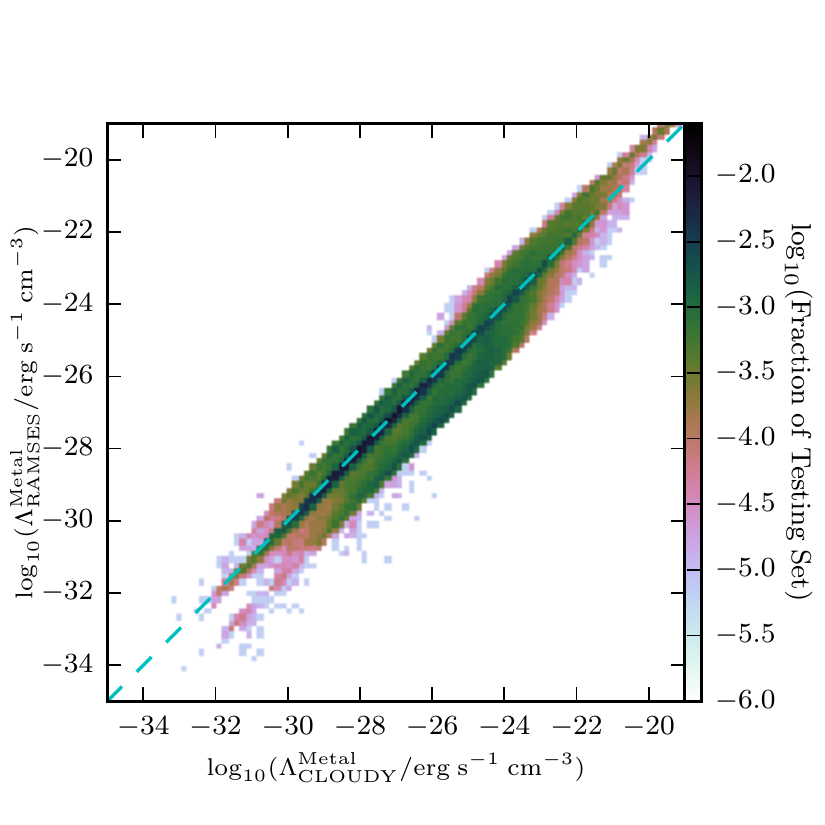}}
\caption{Comparison of the metal-line cooling function currently used in the public version of the {\small RAMSES} code versus the results from our {\small CLOUDY} calculations that include the effects of the local radiation field. The cyan dashed line represents where $\Lambda_{\rm CLOUDY}^{\rm Metal}=\Lambda_{\rm RAMSES}^{\rm Metal}$. For cells with a weak radiation field, the results agree well however there are strong systematic offsets that result from the local radiation source.}
\label{rccomp}
\end{figure}

\begin{figure*}
\centerline{\includegraphics[scale=1,trim={0 0 0 0},clip]{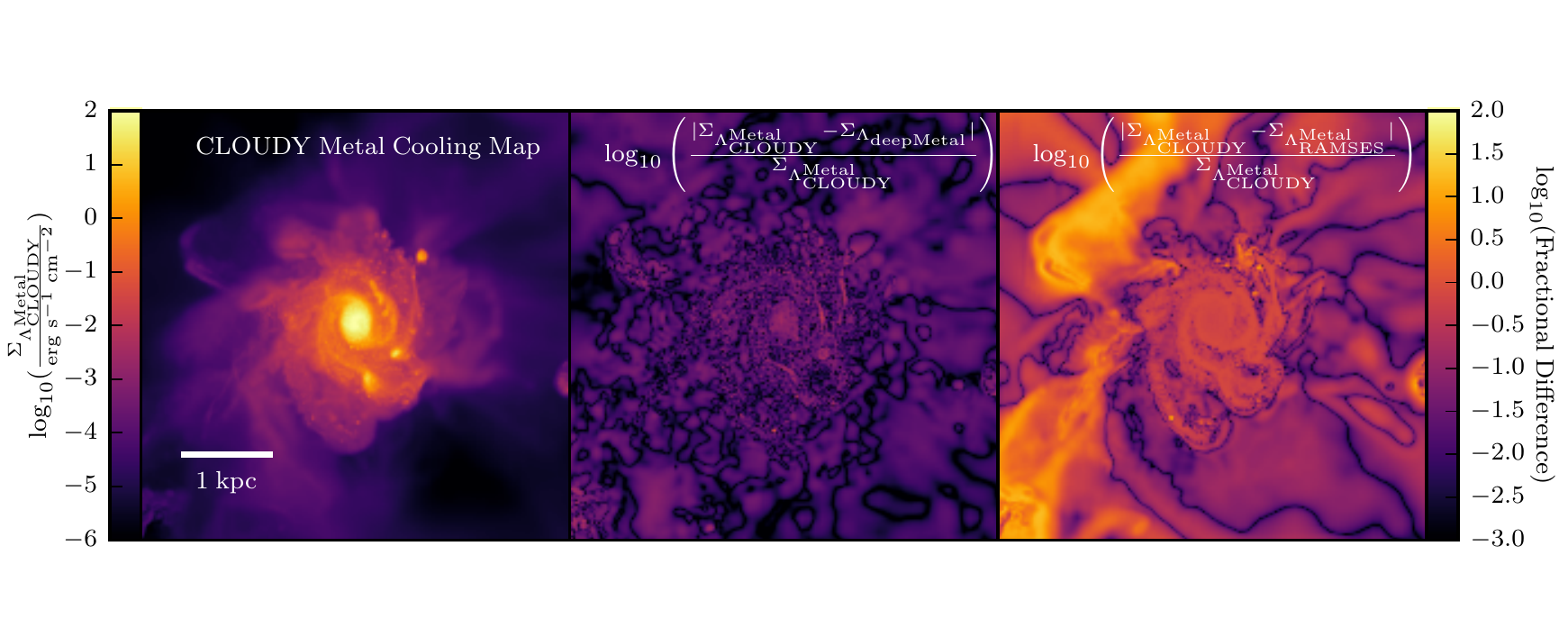}}
\caption{(Left) Map of the metal cooling rate surface density, $\Sigma_{\Lambda_{\rm CLOUDY}^{\rm Metal}}$, computed with {\small CLOUDY} in post-processing in the 5 kpc region around a massive $z=10$ galaxy.  (Centre) Log of the fractional difference between the metal cooling rate surface density measured with {\small deepMetal} and {\small CLOUDY}. (Right) Log of the fractional difference between the metal cooling rate surface density computed with the {\small RAMSES} metal-line cooling function and {\small CLOUDY}.  The fractional difference between {\small deepMetal} and {\small CLOUDY} is extremely small at all locations.  In contrast, there are obvious systematic differences in the regions surrounding the disc when using the {\small RAMSES} metal-line cooling function.  This may impact how gas cools and collapses onto the galaxy.   }
\label{cmcomp}
\end{figure*}

Since it is well established that the local radiation field has a significant effect on the metal-line cooling rates, it is interesting to probe how the metal-line cooling function in the public version of {\small RAMSES} performs compared to our {\small CLOUDY} simulations that take into account the radiation.  In Figure~\ref{rccomp}, we plot a 2D histogram of the cooling rates generated by {\small CLOUDY} against the {\small RAMSES} metal-line cooling function for the cells in our test set.  For cells with a weak radiation field, the {\small RAMSES} metal-line cooling function performs reasonably well as many of the cells fall on the 1-to-1 line in the plot.  The median fractional difference between what is currently being used in {\small RAMSES} is only $\sim 56.5\%$ different from what we predict with our {\small CLOUDY} models.  Part of this difference stems from the fact that a different version of {\small CLOUDY} was used to generate the cooling rates for {\small RAMSES} at $T>10^4$~K where the metallicity dependence is a scaling term, while at low temperatures, {\small RAMSES} employs a fitting function from \cite{Rosen1995}.  Nevertheless, there are still large systematic differences evident in Figure~\ref{rccomp} where the {\small RAMSES} cooling function predicts a cooling rate that is an order of magnitude higher or lower compared to our {\small CLOUDY} models.  If we measure the fractional difference between the two sets of metal-line cooling rates at the 90th and 99th percentile, we find a difference of 1856.5\% and 3360.1\% respectively (compared to 10.5\% and 29.4\% for {\small deepMetal}).  In fact, more than 15\% of the cells in the testing set have metal-line cooling rates that have a fractional difference of more than an order of magnitude.  Hence {\small deepMetal} can provide a significant improvement over what is currently implemented in the code.  Note that the {\small RAMSES} cooling function is not unique in this regard.  Other common simulation suites that employ equilibrium metal-line cooling functions generated with {\small CLOUDY}, even those that use a homogeneous UV background, will fail to capture the effects of the local radiation field and will potentially deviate by more than an order of magnitude from the metal-line cooling rate calculated under the influence of the local radiation source. 

\begin{figure}
\centerline{\includegraphics[scale=1,trim={0 0.3cm 0 1cm},clip]{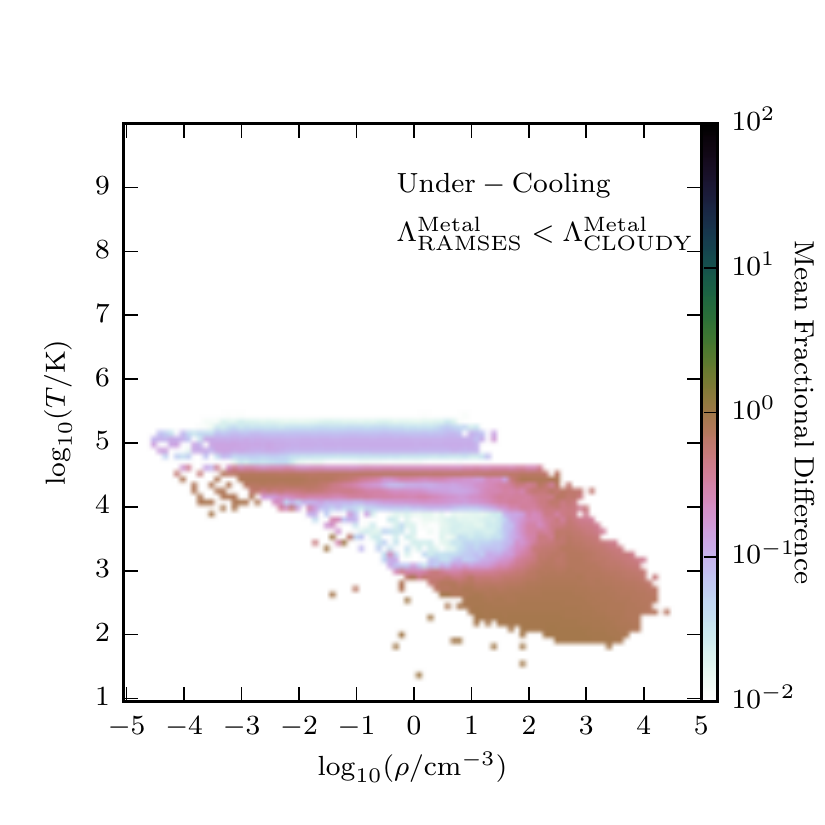}}
\centerline{\includegraphics[scale=1,trim={0 0.3cm 0 1cm},clip]{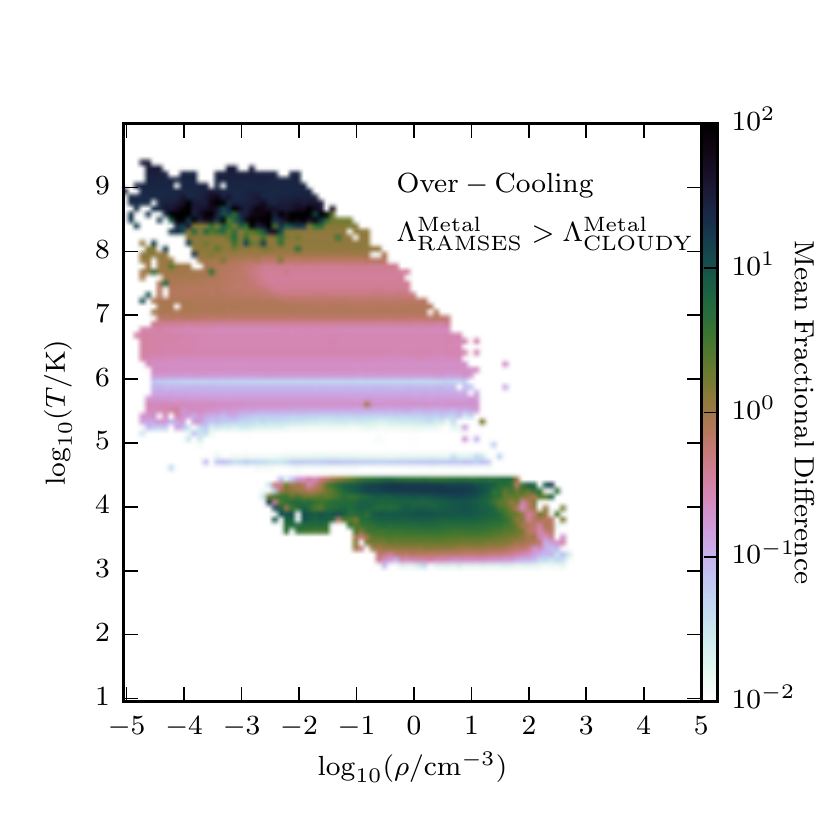}}
\caption{Phase-space diagram of density versus temperature weighted by the mean fractional difference between the {\small RAMSES} metal-line cooling function and the metal-line cooling rates derived with {\small CLOUDY} under the influence of the local radiation field.  The top panel shows the regions where {\small RAMSES} under-predicts the cooling rate (i.e. the gas cools too slowly) while the bottom panel shows where {\small RAMSES} over-predicts the cooling rate (i.e. the gas cools too fast).  The largest differences occur at very high temperatures (where metal-line cooling has no influence) and at $10^4<T/(\rm K)<10^{4.5}$ where metal-line cooling often dominates the total cooling rate.}
\label{pscomp}
\end{figure}

It is important to understand where exactly in the simulation the {\small RAMSES} metal-line cooling function is failing catastrophically compared to the one computed with {\small CLOUDY}. Future simulations are needed to test the impact of this on the baryon cycle and star formation rates. In Figure~\ref{cmcomp}, we show a map of the cooling rate surface density, $\Sigma_{\Lambda}$~$({\rm erg\ s^{-1}\ cm^{-2}})$ as well as the logarithm of the fractional difference of this quantity between what is predicted with {\small deepCool} and the {\small RAMSES} metal-line cooling function.  There are no obvious spatial systematic differences between {\small deepCool} and {\small CLOUDY}.  In contrast, the {\small RAMSES} metal-line cooling function performs worse in the more diffuse gas surrounding the galaxy. The {\small RAMSES} metal-line cooling function tends to cool gas much more efficiently in these regions (see also Figure~\ref{rccomp} and Figure~\ref{pscomp}).  This is particularly important for the baryon cycle in galaxies where feedback is expected to remove some fraction of the baryons from the disk which can then be re-accreted on some time scale dependent on their ability to cool \citep[e.g.][]{Christensen2016}.  A decrease in halo gas cooling was also found in the simulations of \cite{Kannan2014} that aimed to model the effects of local photoionisation feedback on the cooling rates.  If the gas cools much less efficiently due to the presence of a local radiation field, less energetic feedback would be required to prevent the gas from cooling and forming stars. Future work will carefully examine the overall impact of the improved cooling rate calculation on galaxy formation. 

In Figure~\ref{pscomp}, we show 2D phase-space histograms of density versus temperature where we have weighted the cells in the histogram by the log mean fractional error of the cooling rate computed with {\small RAMSES} compared to {\small CLOUDY}.  The top panel shows the results in the regions where {\small RAMSES} under-predicts the cooling rate and the bottom panel shows where {\small RAMSES} over-predicts the cooling rate.  There are certain regions of this phase-space where the old {\small RAMSES} cooling function performs very well and these are observable both as the light bands in Figure~\ref{pscomp} and the dark contours in the right panel of Figure~\ref{cmcomp}.  The largest differences occur at $T>10^8$ K which should not make a large difference as metal-line cooling represents a negligible fraction of the total cooling rate at these temperatures for the metallicity we have considered.  However, there are also substantial errors at $T\sim10^4$ K where metal-lines often dominate the cooling rate.  It is this region of phase-space where we expect that the adopted radiation dependent metal-line cooling rates will make the biggest impact on the simulations.  

\section{Discussion}
\label{discussion}

\subsection{Strategy for Deployment}
Apart from speed of execution and ease of development, one of the main advantages of employing a simple set of cooling tables in cosmological simulations that are temperature, redshift, density, and metallicity dependent (the four common features in most sets of cooling tables), is that no prior knowledge of the simulation is needed before the tables are generated.  In order to develop {\small deepCool}, {\small deepHeat}, and {\small deepMetal}, we first ran a high-resolution simulation with on-the-fly multi-frequency radiation transfer, and used the results of this to train an artificial neural network.  As we stressed earlier, the results from machine learning algorithms are generally only as good as the data that they are trained on.  Thus it is favourable to train the algorithm on simulation data that is as close as possible to where one plans to deploy the trained algorithm.  Since for an expensive simulation, it is often not possible to run them twice, this begs the question of how we intend the algorithm to be used if we must run the simulation to know the cell properties well enough to train the algorithm.

Before launching production runs, multiple other parameters (e.g. those related to feedback) require ``tuning'' in order to calibrate the simulation to reproduce observational properties such as the stellar mass-halo mass function from abundance matching, the stellar mass function at a given redshift, or even the timing of reionization.  It is also common practice to run multiple smaller box simulations at lower and higher resolution in order to understand the convergence properties of the simulations.  It is for these test runs that a representative sample of cells can be extracted and run through codes such as {\small CLOUDY} that can very accurately measure the cooling function for a variety of conditions (i.e. radiation field, cosmic ray background, metal abundance, etc.).  The neural networks can then be trained on this set and deployed for production.  The process of running the {\small CLOUDY} model and training the network is much cheaper computationally compared to an actual state-of-the-art cosmological simulation with the advantage of reducing errors in the cooling rates by more than an order of magnitude.  Hence we believe that the upside of this approach is well worth the small additional computational cost of training the network.

\subsection{Isolated Disk Test}
\begin{figure}
\centerline{\includegraphics[scale=1,trim={0 0 0 0},clip]{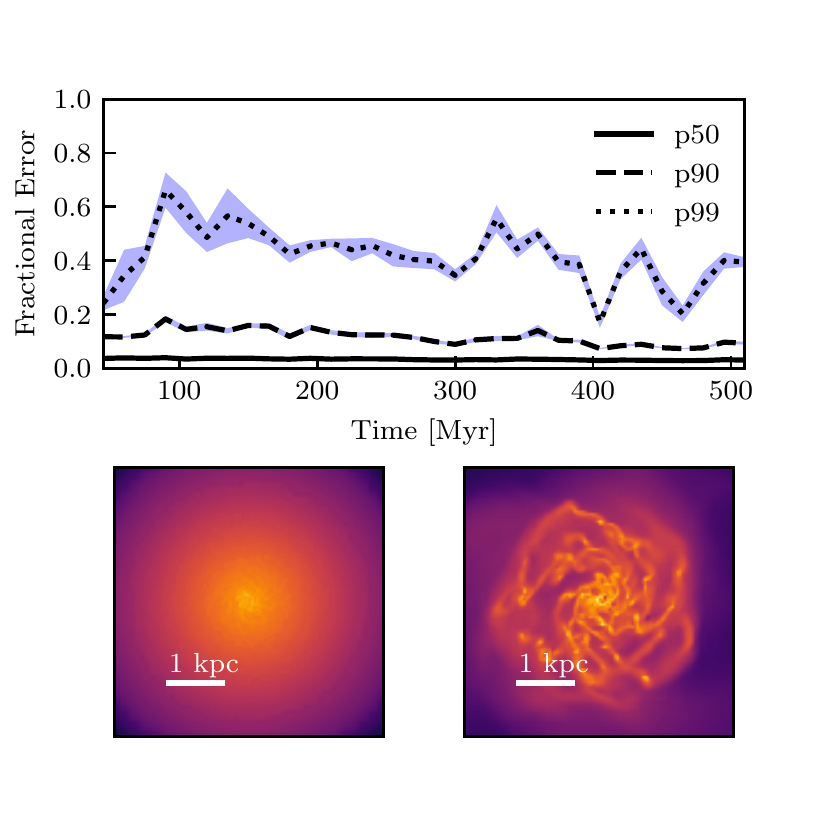}}
\caption{(Top.) Fractional difference $\frac{|\Lambda_{\rm deepMetal}-\Lambda_{\rm CLOUDY}^{\rm Metal}|}{\Lambda_{\rm CLOUDY}^{\rm Metal}}$ between the predicted metal-line cooling rates and the true cooling rates generated by {\rm CLOUDY} as a function of time in the isolated disk test.  The solid, dashed, and dotted lines represent the 50th, 90th, and 99th percentile in the fractional difference.  The blue bands represent error bars on these fractional differences generated by partitioning each set of 10,000 cells into ten subsets and calculating the standard deviation among the fractional differences in each subset.  (Bottom.) Face-on images of the gas surface density in the second ($t=15$ Myr) and final ($t=500$ Myr) simulation snapshots.}
\label{diskplot}
\end{figure}

We seek to understand the computational cost and accuracy of the new algorithm, and deploying it in a real simulation makes for an ideal test.  For this reason, we have set up a small isolated disk galaxy (G8) with a halo mass of $10^{10}\,$M$_{\odot}$ and gas mass of $3.5\times10^8\,$M$_{\odot}$ (see \citealt{Rosdahl2015b} for details).  The simulation uses very similar physics to the run described in \S\ref{sim_details}; however, we have reduced the number of radiation bins to three (HI ionising: 13.6 -- 24.59 eV, HeI ionising: 24.59 -- 54.42 eV, and HeII ionising: $>54.42$ eV) to both reduce the computational cost of the simulation and to run something that is more representative of standard cosmological radiation hydrodynamics simulations that model reionization.  We allow the simulation to refine to a maximum physical resolution of 20 pc.  In the first instance, the simulation is evolved for 500 Myr, producing simulation outputs approximately every 15 Myr.  We then select a small subset of cells in the second output (the output after the first star has formed in the simulation where most of the gas is self-shielded) as well as the final output at 500 Myr (when the galaxy has evolved to different parts of temperature-density phase-space and metal enrichment has occurred) to train {\small deepMetal}.  Face-on images of the gas surface density for these outputs are shown in the bottom row of Figure~\ref{diskplot}.  Training on both outputs allows our algorithm to encapsulate a wide region of parameter space.  

To determine how well we expect {\small deepMetal} to perform in terms of accuracy, we have selected a random subset of 10,000 gas cells in all intermediate simulation outputs (between 50 and 500 Myr) and run {\small CLOUDY} to calculate the cooling rates.  We then used {\small deepMetal} to predict the cooling rates not only for cells on which it has not been trained, but also for cells in completely different simulation snapshots.  In the top row of Figure~\ref{diskplot} we show the median, 90th and 99th percentile fractional error in the {\small deepMetal} predictions and we find that it performs very well with median errors of only a few percent.  This once again demonstrates that the algorithm is extremely generalisable.  

Finally, we rerun the simulation for 500 Myr replacing the standard {\small RAMSES} metal-line cooling module with {\small deepMetal} to calculate any changes in runtime.  In terms of computational time to complete the calculation, we find little to no increase in the simulation run time for the simulation that uses {\small deepMetal} compared to that which used the standard {\small RAMSES} cooling function.  Although the {\small deepMetal} subroutine is approximately four times slower than the standard {\small RAMSES} cooling function due to additional matrix operations, cooling represents only a few percent of the total CPU time; hence there is no significant computational loss from using {\small deepCool}.  In future work, we will systematically examine the effects of using this new cooling routine on the star formation history, and evolution of gas in simulated galaxies. 

\subsection{Alternative Uses}
In this work, we focused on measuring the cooling and heating rates of astrophysical gases by training an artificial neural network to replace a much more expensive photoionisation code ({\small CLOUDY}).  These types of codes are used for a variety of other purposes and hence it is safe to assume that neural networks may be able to replace their functionality for other problems.  For example, Katz~et~al.~2018~(submitted) first employed this algorithm to measure emission line properties of high-redshift galaxies (i.e. [CII]~157.6$\mu$m, [OIII]~88.33$\mu$m, H$\alpha$, H$\beta$, etc.) in order to make predictions for ALMA and JWST.  {\small CLOUDY} was run on the same $\sim850,000$ cells used in this work to accurately measure the luminosities of each emission line for each cell.  A random forest was then trained on this data and used to accurately predict the emission line properties of thousands of galaxies in the simulation.  This type of algorithm can be similarly trained and used, for example, to quickly estimate ionisation states of metals for absorption line studies, nebular emission lines to make mock spectral energy distributions, or to measure molecular properties of gases to determine, for example, H$_2$ content.  These types of problems represent an ideal use case for artificial neural networks in the context of astrophysics.

\section{Conclusions}
\label{conclusions}
In this work, we have demonstrated that artificial neural networks are an ideal tool for accurately estimating cooling rates, heating rates, and metal-line cooling rates of irradiated gas.  Because of their speed of execution and low memory cost, they represent an efficient class of algorithms that can be deployed in state-of-the-art cosmological simulations to improve the modelling of radiative cooling, and probably a variety of other physical processes.  We have used the results of a high-resolution cosmological radiation hydrodynamics simulation to train neural networks to measure cooling rates, heating rates, and metal-line cooling rates that are dependent on a local radiation field and our results can be summarised as follows:

\begin{itemize}
\item Our trained artificial neural networks, {\small deepCool}, {\small deepHeat}, and {\small deepMetal}, can accurately predict the total cooling, total heating, and metal-line cooling rates of irradiated gas to accuracies of $6.3\%$, $6.6\%$, and $3.8\%$, respectively, at the 50th percentile.  For metal-line cooling rates, {\small deepMetal} can predict 99\% of all cells to within a 30\% accuracy.

\item  {\small deepMetal} can become inaccurate at $T>10^8$ K; however, in this temperature range, the contribution from metal-line cooling represents only $0.01\%$ of the total cooling rate indicating that the uncertainty in our model will have no impact on the gas thermochemistry.  Similarly, {\small deepMetal} can become inaccurate at $T\sim100$ K, where very few cells in the simulation were available to train the algorithm.  These cells are expected to be rapidly forming stars meaning that these inaccuracies should last for only a very short period of time.

\item Standard CIE cooling functions, are particularly error prone compared to the irradiated gas cooling function at $T\sim10^4$ K with densities $-2\lesssim\log_{10}(\rho/{\rm cm^{-3}})\lesssim3$ where differences in the cooling rate can be more than an order of magnitude.  Neglecting the impact of radiation on the cooling rates means that more than 15\% of the cells in the simulation will be subject to a cooling rate that is an order of magnitude too strong.  This is particularly important for the gas in and around the dense centre of the galaxy and could impact the baryon cycle.

\end{itemize}

It remains to be determined exactly how much the star formation rate of a galaxy is altered when switching between radiation-independent and radiation dependent cooling rates in a full cosmological simulation that models the full baryon cycle.  We have demonstrated that the impact of the local radiation field on the cooling rates can be strong and the effect may be even greater in the context of an AGN \citep{Gnedin2012}.  Given the computational efficiency and accuracy of our method, we suggest that future simulations take into account the effects of radiation (and other physics, such as non-solar abundances and cosmic rays) on the cooling rates using similar tools that are easily embeddable into modern simulation codes.

\section*{Acknowledgements}
This work made considerable use of the open source analysis software {\small PYNBODY} \citep{pynbody}. TPG thanks Brasenose College. HK thanks Brasenose College and the support of the Nicholas Kurti Junior Fellowship as well as the Beecroft Fellowship. TK was supported by the National Research Foundation of Korea (No. 2017R1A5A1070354 and No. 2018036146). JB and JR acknowledge support from the ORAGE project from the Agence Nationale de la Recherche under grant
ANR-14-CE33-0016-03.

This work was performed using the DiRAC/Darwin Supercomputer hosted by the University of Cambridge High Performance Computing Service (http://www.hpc.cam.ac.uk/), provided by Dell Inc. using Strategic Research Infrastructure Funding from the Higher Education Funding Council for England and funding from the Science and Technology Facilities Council. 

This work used the DiRAC Complexity system, operated by the University of Leicester IT Services, which forms part of the STFC DiRAC HPC Facility (www.dirac.ac.uk). This equipment is funded by BIS National E-Infrastructure capital grant ST/K000373/1 and  STFC DiRAC Operations grant ST/K0003259/1. 

Furthermore, this work used the DiRAC Data Centric system at Durham University, operated by the Institute for Computational Cosmology on behalf of the STFC DiRAC HPC Facility (www.dirac.ac.uk).  This equipment was funded by the BIS National E-infrastructure capital grant ST/K00042X/1, STFC capital grant ST/K00087X/1, DiRAC operations grant ST/K003267/1 and Durham University.  DiRAC is part of the National E-Infrastructure.




\bibliographystyle{mnras}
\bibliography{mnras} 

\begin{thebibliography}{}
\makeatletter
\relax
\def\mn@urlcharsother{\let\do\@makeother \do\$\do\&\do\#\do\^\do\_\do\%\do\~}
\def\mn@doi{\begingroup\mn@urlcharsother \@ifnextchar [ {\mn@doi@}
  {\mn@doi@[]}}
\def\mn@doi@[#1]#2{\def\@tempa{#1}\ifx\@tempa\@empty \href
  {http://dx.doi.org/#2} {doi:#2}\else \href {http://dx.doi.org/#2} {#1}\fi
  \endgroup}
\def\mn@eprint#1#2{\mn@eprint@#1:#2::\@nil}
\def\mn@eprint@arXiv#1{\href {http://arxiv.org/abs/#1} {{\tt arXiv:#1}}}
\def\mn@eprint@dblp#1{\href {http://dblp.uni-trier.de/rec/bibtex/#1.xml}
  {dblp:#1}}
\def\mn@eprint@#1:#2:#3:#4\@nil{\def\@tempa {#1}\def\@tempb {#2}\def\@tempc
  {#3}\ifx \@tempc \@empty \let \@tempc \@tempb \let \@tempb \@tempa \fi \ifx
  \@tempb \@empty \def\@tempb {arXiv}\fi \@ifundefined
  {mn@eprint@\@tempb}{\@tempb:\@tempc}{\expandafter \expandafter \csname
  mn@eprint@\@tempb\endcsname \expandafter{\@tempc}}}

\bibitem[\protect\citeauthoryear{{Bakes} \& {Tielens}}{{Bakes} \&
  {Tielens}}{1994}]{Bakes1994}
{Bakes} E.~L.~O.,  {Tielens} A.~G.~G.~M.,  1994, \mn@doi [\apj]
  {10.1086/174188}, \href {http://adsabs.harvard.edu/abs/1994ApJ...427..822B}
  {427, 822}

\bibitem[\protect\citeauthoryear{{Black}}{{Black}}{1987}]{Black1987}
{Black} J.~H.,  1987, in {Hollenbach} D.~J.,  {Thronson} Jr. H.~A.,  eds,
  Astrophysics and Space Science Library Vol. 134, Interstellar Processes. pp
  731--744, \mn@doi{10.1007/978-94-009-3861-8_27}

\bibitem[\protect\citeauthoryear{Breiman}{Breiman}{2001}]{Breiman2001}
Breiman L.,  2001, \mn@doi [Mach. Learn.] {10.1023/A:1010933404324}, 45, 5

\bibitem[\protect\citeauthoryear{{Cantalupo}}{{Cantalupo}}{2010}]{Cantalupo2010}
{Cantalupo} S.,  2010, \mn@doi [\mnras] {10.1111/j.1745-3933.2010.00806.x},
  \href {http://adsabs.harvard.edu/abs/2010MNRAS.403L..16C} {403, L16}

\bibitem[\protect\citeauthoryear{{Christensen}, {Dav{\'e}}, {Governato},
  {Pontzen}, {Brooks}, {Munshi}, {Quinn}  \& {Wadsley}}{{Christensen}
  et~al.}{2016}]{Christensen2016}
{Christensen} C.~R.,  {Dav{\'e}} R.,  {Governato} F.,  {Pontzen} A.,  {Brooks}
  A.,  {Munshi} F.,  {Quinn} T.,   {Wadsley} J.,  2016, \mn@doi [\apj]
  {10.3847/0004-637X/824/1/57}, \href
  {http://adsabs.harvard.edu/abs/2016ApJ...824...57C} {824, 57}

\bibitem[\protect\citeauthoryear{{Cox} \& {Daltabuit}}{{Cox} \&
  {Daltabuit}}{1971}]{Cox1971}
{Cox} D.~P.,  {Daltabuit} E.,  1971, \mn@doi [\apj] {10.1086/151009}, \href
  {http://adsabs.harvard.edu/abs/1971ApJ...167..113C} {167, 113}

\bibitem[\protect\citeauthoryear{{Cox} \& {Tucker}}{{Cox} \&
  {Tucker}}{1969}]{Cox1969}
{Cox} D.~P.,  {Tucker} W.~H.,  1969, \mn@doi [\apj] {10.1086/150144}, \href
  {http://adsabs.harvard.edu/abs/1969ApJ...157.1157C} {157, 1157}

\bibitem[\protect\citeauthoryear{{Crain} et~al.,}{{Crain}
  et~al.}{2015}]{Crain2015}
{Crain} R.~A.,  et~al., 2015, \mn@doi [\mnras] {10.1093/mnras/stv725}, \href
  {http://adsabs.harvard.edu/abs/2015MNRAS.450.1937C} {450, 1937}

\bibitem[\protect\citeauthoryear{{Dalgarno} \& {McCray}}{{Dalgarno} \&
  {McCray}}{1972}]{Dalgarno1972}
{Dalgarno} A.,  {McCray} R.~A.,  1972, \mn@doi [\araa]
  {10.1146/annurev.aa.10.090172.002111}, \href
  {http://adsabs.harvard.edu/abs/1972ARA%26A..10..375D} {10, 375}

\bibitem[\protect\citeauthoryear{{Dolag}, {Bykov}  \& {Diaferio}}{{Dolag}
  et~al.}{2008}]{Dolag2008}
{Dolag} K.,  {Bykov} A.~M.,   {Diaferio} A.,  2008, \mn@doi [\ssr]
  {10.1007/s11214-008-9319-2}, \href
  {http://adsabs.harvard.edu/abs/2008SSRv..134..311D} {134, 311}

\bibitem[\protect\citeauthoryear{{Dubois} et~al.,}{{Dubois}
  et~al.}{2014}]{Dubois2014}
{Dubois} Y.,  et~al., 2014, \mn@doi [\mnras] {10.1093/mnras/stu1227}, \href
  {http://adsabs.harvard.edu/abs/2014MNRAS.444.1453D} {444, 1453}

\bibitem[\protect\citeauthoryear{{Efstathiou}}{{Efstathiou}}{1992}]{Efstathiou1992}
{Efstathiou} G.,  1992, \mn@doi [\mnras] {10.1093/mnras/256.1.43P}, \href
  {http://adsabs.harvard.edu/abs/1992MNRAS.256P..43E} {256, 43P}

\bibitem[\protect\citeauthoryear{{Eldridge}, {Izzard}  \& {Tout}}{{Eldridge}
  et~al.}{2008}]{Eldridge2008}
{Eldridge} J.~J.,  {Izzard} R.~G.,   {Tout} C.~A.,  2008, \mn@doi [\mnras]
  {10.1111/j.1365-2966.2007.12738.x}, \href
  {http://adsabs.harvard.edu/abs/2008MNRAS.384.1109E} {384, 1109}

\bibitem[\protect\citeauthoryear{{Ferland}, {Korista}, {Verner}, {Ferguson},
  {Kingdon}  \& {Verner}}{{Ferland} et~al.}{1998}]{Ferland1998}
{Ferland} G.~J.,  {Korista} K.~T.,  {Verner} D.~A.,  {Ferguson} J.~W.,
  {Kingdon} J.~B.,   {Verner} E.~M.,  1998, \mn@doi [\pasp] {10.1086/316190},
  \href {http://adsabs.harvard.edu/abs/1998PASP..110..761F} {110, 761}

\bibitem[\protect\citeauthoryear{{Ferland} et~al.,}{{Ferland}
  et~al.}{2017}]{Ferland2017}
{Ferland} G.~J.,  et~al., 2017, \rmxaa, \href
  {http://adsabs.harvard.edu/abs/2017RMxAA..53..385F} {53, 385}

\bibitem[\protect\citeauthoryear{Glorot, Bordes  \& Bengio}{Glorot
  et~al.}{2011}]{relu}
Glorot X.,  Bordes A.,   Bengio Y.,  2011, in Gordon G.,  Dunson D.,
  Dud\'{i}k M.,  eds,  Proceedings of Machine Learning Research Vol. 15,
  Proceedings of the Fourteenth International Conference on Artificial
  Intelligence and Statistics. PMLR, Fort Lauderdale, FL, USA, pp 315--323,
  \url {http://proceedings.mlr.press/v15/glorot11a.html}

\bibitem[\protect\citeauthoryear{{Gnat} \& {Ferland}}{{Gnat} \&
  {Ferland}}{2012}]{Gnat2012}
{Gnat} O.,  {Ferland} G.~J.,  2012, \mn@doi [\apjs]
  {10.1088/0067-0049/199/1/20}, \href
  {http://adsabs.harvard.edu/abs/2012ApJS..199...20G} {199, 20}

\bibitem[\protect\citeauthoryear{{Gnedin}}{{Gnedin}}{2014}]{Gnedin2014}
{Gnedin} N.~Y.,  2014, \mn@doi [\apj] {10.1088/0004-637X/793/1/29}, \href
  {http://adsabs.harvard.edu/abs/2014ApJ...793...29G} {793, 29}

\bibitem[\protect\citeauthoryear{{Gnedin} \& {Hollon}}{{Gnedin} \&
  {Hollon}}{2012}]{Gnedin2012}
{Gnedin} N.~Y.,  {Hollon} N.,  2012, \mn@doi [\apjs]
  {10.1088/0067-0049/202/2/13}, \href
  {http://adsabs.harvard.edu/abs/2012ApJS..202...13G} {202, 13}

\bibitem[\protect\citeauthoryear{{Grevesse}, {Asplund}, {Sauval}  \&
  {Scott}}{{Grevesse} et~al.}{2010}]{Grevesse2010}
{Grevesse} N.,  {Asplund} M.,  {Sauval} A.~J.,   {Scott} P.,  2010, \mn@doi
  [\apss] {10.1007/s10509-010-0288-z}, \href
  {http://adsabs.harvard.edu/abs/2010Ap%26SS.328..179G} {328, 179}

\bibitem[\protect\citeauthoryear{{Haardt} \& {Madau}}{{Haardt} \&
  {Madau}}{1996}]{Haardt1996}
{Haardt} F.,  {Madau} P.,  1996, \mn@doi [\apj] {10.1086/177035}, \href
  {http://adsabs.harvard.edu/abs/1996ApJ...461...20H} {461, 20}

\bibitem[\protect\citeauthoryear{{Haardt} \& {Madau}}{{Haardt} \&
  {Madau}}{2012}]{Haardt2012}
{Haardt} F.,  {Madau} P.,  2012, \mn@doi [\apj] {10.1088/0004-637X/746/2/125},
  \href {http://adsabs.harvard.edu/abs/2012ApJ...746..125H} {746, 125}

\bibitem[\protect\citeauthoryear{{Hahn} \& {Abel}}{{Hahn} \&
  {Abel}}{2011}]{Hahn2011}
{Hahn} O.,  {Abel} T.,  2011, \mn@doi [\mnras]
  {10.1111/j.1365-2966.2011.18820.x}, \href
  {http://adsabs.harvard.edu/abs/2011MNRAS.415.2101H} {415, 2101}

\bibitem[\protect\citeauthoryear{Ho}{Ho}{1995}]{Ho1995}
Ho T.~K.,  1995, in Proceedings of the Third International Conference on
  Document Analysis and Recognition (Volume 1) - Volume 1. ICDAR '95.
IEEE Computer Society, Washington, DC, USA, pp 278--, \url
  {http://dl.acm.org/citation.cfm?id=844379.844681}

\bibitem[\protect\citeauthoryear{{Kannan} et~al.,}{{Kannan}
  et~al.}{2014}]{Kannan2014}
{Kannan} R.,  et~al., 2014, \mn@doi [\mnras] {10.1093/mnras/stt2098}, \href
  {http://adsabs.harvard.edu/abs/2014MNRAS.437.2882K} {437, 2882}

\bibitem[\protect\citeauthoryear{{Kannan}, {Vogelsberger}, {Stinson},
  {Hennawi}, {Marinacci}, {Springel}  \& {Macci{\`o}}}{{Kannan}
  et~al.}{2016}]{Kannan2016}
{Kannan} R.,  {Vogelsberger} M.,  {Stinson} G.~S.,  {Hennawi} J.~F.,
  {Marinacci} F.,  {Springel} V.,   {Macci{\`o}} A.~V.,  2016, \mn@doi [\mnras]
  {10.1093/mnras/stw463}, \href
  {http://adsabs.harvard.edu/abs/2016MNRAS.458.2516K} {458, 2516}

\bibitem[\protect\citeauthoryear{{Katz}, {Weinberg}, {Hernquist}  \&
  {Miralda-Escude}}{{Katz} et~al.}{1996}]{Katz1996}
{Katz} N.,  {Weinberg} D.~H.,  {Hernquist} L.,   {Miralda-Escude} J.,  1996,
  \mn@doi [\apjl] {10.1086/309900}, \href
  {http://adsabs.harvard.edu/abs/1996ApJ...457L..57K} {457, L57}

\bibitem[\protect\citeauthoryear{{Katz}, {Kimm}, {Sijacki}  \&
  {Haehnelt}}{{Katz} et~al.}{2017}]{Katz2017}
{Katz} H.,  {Kimm} T.,  {Sijacki} D.,   {Haehnelt} M.~G.,  2017, \mn@doi
  [\mnras] {10.1093/mnras/stx608}, \href
  {http://adsabs.harvard.edu/abs/2017MNRAS.468.4831K} {468, 4831}

\bibitem[\protect\citeauthoryear{{Katz}, {Laporte}, {Ellis}, {Devriendt}  \&
  {Slyz}}{{Katz} et~al.}{2018b}]{Katz2018}
{Katz} H.,  {Laporte} N.,  {Ellis} R.~S.,  {Devriendt} J.,   {Slyz} A.,  2018b,
  preprint, \href {http://adsabs.harvard.edu/abs/2018arXiv180907210K} {}
  (\mn@eprint {arXiv} {1809.07210})

\bibitem[\protect\citeauthoryear{{Katz}, {Kimm}, {Haehnelt}, {Sijacki},
  {Rosdahl}  \& {Blaizot}}{{Katz} et~al.}{2018a}]{Katz2018b}
{Katz} H.,  {Kimm} T.,  {Haehnelt} M.~G.,  {Sijacki} D.,  {Rosdahl} J.,
  {Blaizot} J.,  2018a, preprint, \href
  {http://adsabs.harvard.edu/abs/2018arXiv180607474K} {} (\mn@eprint {arXiv}
  {1806.07474})

\bibitem[\protect\citeauthoryear{{Katz}, {Kimm}, {Haehnelt}, {Sijacki},
  {Rosdahl}  \& {Blaizot}}{{Katz} et~al.}{2018c}]{Katz2018a}
{Katz} H.,  {Kimm} T.,  {Haehnelt} M.,  {Sijacki} D.,  {Rosdahl} J.,
  {Blaizot} J.,  2018c, \mn@doi [\mnras] {10.1093/mnras/sty1225}, \href
  {http://adsabs.harvard.edu/abs/2018MNRAS.478.4986K} {478, 4986}

\bibitem[\protect\citeauthoryear{{Kimm}, {Cen}, {Devriendt}, {Dubois}  \&
  {Slyz}}{{Kimm} et~al.}{2015}]{Kimm2015}
{Kimm} T.,  {Cen} R.,  {Devriendt} J.,  {Dubois} Y.,   {Slyz} A.,  2015,
  \mn@doi [\mnras] {10.1093/mnras/stv1211}, \href
  {http://adsabs.harvard.edu/abs/2015MNRAS.451.2900K} {451, 2900}

\bibitem[\protect\citeauthoryear{{Kimm}, {Katz}, {Haehnelt}, {Rosdahl},
  {Devriendt}  \& {Slyz}}{{Kimm} et~al.}{2017}]{Kimm2017}
{Kimm} T.,  {Katz} H.,  {Haehnelt} M.,  {Rosdahl} J.,  {Devriendt} J.,   {Slyz}
  A.,  2017, \mn@doi [\mnras] {10.1093/mnras/stx052}, \href
  {http://adsabs.harvard.edu/abs/2017MNRAS.466.4826K} {466, 4826}

\bibitem[\protect\citeauthoryear{Kingma \& Ba}{Kingma \& Ba}{2014}]{adam}
Kingma D.~P.,  Ba J.,  2014, CoRR, abs/1412.6980

\bibitem[\protect\citeauthoryear{{Kobayashi}, {Umeda}, {Nomoto}, {Tominaga}  \&
  {Ohkubo}}{{Kobayashi} et~al.}{2006}]{Kobayashi2006}
{Kobayashi} C.,  {Umeda} H.,  {Nomoto} K.,  {Tominaga} N.,   {Ohkubo} T.,
  2006, \mn@doi [\apj] {10.1086/508914}, \href
  {http://adsabs.harvard.edu/abs/2006ApJ...653.1145K} {653, 1145}

\bibitem[\protect\citeauthoryear{{Kroupa}}{{Kroupa}}{2001}]{Kroupa2001}
{Kroupa} P.,  2001, \mn@doi [\mnras] {10.1046/j.1365-8711.2001.04022.x}, \href
  {http://adsabs.harvard.edu/abs/2001MNRAS.322..231K} {322, 231}

\bibitem[\protect\citeauthoryear{{Levermore}}{{Levermore}}{1984}]{Levermore1984}
{Levermore} C.~D.,  1984, \mn@doi [\jqsrt] {10.1016/0022-4073(84)90112-2},
  \href {http://adsabs.harvard.edu/abs/1984JQSRT..31..149L} {31, 149}

\bibitem[\protect\citeauthoryear{{Nomoto}, {Hashimoto}, {Tsujimoto},
  {Thielemann}, {Kishimoto}, {Kubo}  \& {Nakasato}}{{Nomoto}
  et~al.}{1997a}]{Nomoto1997b}
{Nomoto} K.,  {Hashimoto} M.,  {Tsujimoto} T.,  {Thielemann} F.-K.,
  {Kishimoto} N.,  {Kubo} Y.,   {Nakasato} N.,  1997a, \mn@doi [Nuclear Physics
  A] {10.1016/S0375-9474(97)00076-6}, \href
  {http://adsabs.harvard.edu/abs/1997NuPhA.616...79N} {616, 79}

\bibitem[\protect\citeauthoryear{{Nomoto}, {Iwamoto}, {Nakasato}, {Thielemann},
  {Brachwitz}, {Tsujimoto}, {Kubo}  \& {Kishimoto}}{{Nomoto}
  et~al.}{1997b}]{Nomoto1997a}
{Nomoto} K.,  {Iwamoto} K.,  {Nakasato} N.,  {Thielemann} F.-K.,  {Brachwitz}
  F.,  {Tsujimoto} T.,  {Kubo} Y.,   {Kishimoto} N.,  1997b, \mn@doi [Nuclear
  Physics A] {10.1016/S0375-9474(97)00291-1}, \href
  {http://adsabs.harvard.edu/abs/1997NuPhA.621..467N} {621, 467}

\bibitem[\protect\citeauthoryear{{O'Shea}, {Wise}, {Xu}  \& {Norman}}{{O'Shea}
  et~al.}{2015}]{Oshea2015}
{O'Shea} B.~W.,  {Wise} J.~H.,  {Xu} H.,   {Norman} M.~L.,  2015, \mn@doi
  [\apjl] {10.1088/2041-8205/807/1/L12}, \href
  {http://adsabs.harvard.edu/abs/2015ApJ...807L..12O} {807, L12}

\bibitem[\protect\citeauthoryear{{Pawlik}, {Rahmati}, {Schaye}, {Jeon}  \&
  {Dalla Vecchia}}{{Pawlik} et~al.}{2017}]{Pawlik2017}
{Pawlik} A.~H.,  {Rahmati} A.,  {Schaye} J.,  {Jeon} M.,   {Dalla Vecchia} C.,
  2017, \mn@doi [\mnras] {10.1093/mnras/stw2869}, \href
  {http://adsabs.harvard.edu/abs/2017MNRAS.466..960P} {466, 960}

\bibitem[\protect\citeauthoryear{{Planck Collaboration} et~al.,}{{Planck
  Collaboration} et~al.}{2016}]{Planck2016}
{Planck Collaboration} et~al., 2016, \mn@doi [\aap]
  {10.1051/0004-6361/201525830}, \href
  {http://adsabs.harvard.edu/abs/2016A%26A...594A..13P} {594, A13}

\bibitem[\protect\citeauthoryear{{Pontzen}, {Ro{\v s}kar}, {Stinson}, {Woods},
  {Reed}, {Coles}  \& {Quinn}}{{Pontzen} et~al.}{2013}]{pynbody}
{Pontzen} A.,  {Ro{\v s}kar} R.,  {Stinson} G.~S.,  {Woods} R.,  {Reed} D.~M.,
  {Coles} J.,   {Quinn} T.~R.,  2013, {pynbody: Astrophysics Simulation
  Analysis for Python}

\bibitem[\protect\citeauthoryear{{Raymond}, {Cox}  \& {Smith}}{{Raymond}
  et~al.}{1976}]{Raymond1976}
{Raymond} J.~C.,  {Cox} D.~P.,   {Smith} B.~W.,  1976, \mn@doi [\apj]
  {10.1086/154170}, \href {http://adsabs.harvard.edu/abs/1976ApJ...204..290R}
  {204, 290}

\bibitem[\protect\citeauthoryear{{Richings}, {Schaye}  \&
  {Oppenheimer}}{{Richings} et~al.}{2014a}]{Richings2014a}
{Richings} A.~J.,  {Schaye} J.,   {Oppenheimer} B.~D.,  2014a, \mn@doi [\mnras]
  {10.1093/mnras/stu525}, \href
  {http://adsabs.harvard.edu/abs/2014MNRAS.440.3349R} {440, 3349}

\bibitem[\protect\citeauthoryear{{Richings}, {Schaye}  \&
  {Oppenheimer}}{{Richings} et~al.}{2014b}]{Richings2014b}
{Richings} A.~J.,  {Schaye} J.,   {Oppenheimer} B.~D.,  2014b, \mn@doi [\mnras]
  {10.1093/mnras/stu1046}, \href
  {http://adsabs.harvard.edu/abs/2014MNRAS.442.2780R} {442, 2780}

\bibitem[\protect\citeauthoryear{{Rosdahl}, {Blaizot}, {Aubert}, {Stranex}  \&
  {Teyssier}}{{Rosdahl} et~al.}{2013}]{Rosdahl2013}
{Rosdahl} J.,  {Blaizot} J.,  {Aubert} D.,  {Stranex} T.,   {Teyssier} R.,
  2013, \mn@doi [\mnras] {10.1093/mnras/stt1722}, \href
  {http://adsabs.harvard.edu/abs/2013MNRAS.436.2188R} {436, 2188}

\bibitem[\protect\citeauthoryear{{Rosdahl}, {Schaye}, {Teyssier}  \&
  {Agertz}}{{Rosdahl} et~al.}{2015a}]{Rosdahl2015}
{Rosdahl} J.,  {Schaye} J.,  {Teyssier} R.,   {Agertz} O.,  2015a, \mn@doi
  [\mnras] {10.1093/mnras/stv937}, \href
  {http://adsabs.harvard.edu/abs/2015MNRAS.451...34R} {451, 34}

\bibitem[\protect\citeauthoryear{{Rosdahl}, {Schaye}, {Teyssier}  \&
  {Agertz}}{{Rosdahl} et~al.}{2015b}]{Rosdahl2015b}
{Rosdahl} J.,  {Schaye} J.,  {Teyssier} R.,   {Agertz} O.,  2015b, \mn@doi
  [\mnras] {10.1093/mnras/stv937}, \href
  {http://adsabs.harvard.edu/abs/2015MNRAS.451...34R} {451, 34}

\bibitem[\protect\citeauthoryear{{Rosdahl} et~al.,}{{Rosdahl}
  et~al.}{2018}]{Rosdahl2018}
{Rosdahl} J.,  et~al., 2018, preprint, \href
  {http://adsabs.harvard.edu/abs/2018arXiv180107259R} {} (\mn@eprint {arXiv}
  {1801.07259})

\bibitem[\protect\citeauthoryear{{Rosen} \& {Bregman}}{{Rosen} \&
  {Bregman}}{1995}]{Rosen1995}
{Rosen} A.,  {Bregman} J.~N.,  1995, \mn@doi [\apj] {10.1086/175303}, \href
  {http://adsabs.harvard.edu/abs/1995ApJ...440..634R} {440, 634}

\bibitem[\protect\citeauthoryear{Schmidhuber}{Schmidhuber}{2015}]{ANNs}
Schmidhuber J.,  2015, \mn@doi [Neural Networks]
  {https://doi.org/10.1016/j.neunet.2014.09.003}, 61, 85

\bibitem[\protect\citeauthoryear{{Schmidt}}{{Schmidt}}{1959}]{Schmidt1959}
{Schmidt} M.,  1959, \mn@doi [\apj] {10.1086/146614}, \href
  {http://adsabs.harvard.edu/abs/1959ApJ...129..243S} {129, 243}

\bibitem[\protect\citeauthoryear{{Shen}, {Wadsley}  \& {Stinson}}{{Shen}
  et~al.}{2010}]{Shen2010}
{Shen} S.,  {Wadsley} J.,   {Stinson} G.,  2010, \mn@doi [\mnras]
  {10.1111/j.1365-2966.2010.17047.x}, \href
  {http://adsabs.harvard.edu/abs/2010MNRAS.407.1581S} {407, 1581}

\bibitem[\protect\citeauthoryear{{Shull} \& {van Steenberg}}{{Shull} \& {van
  Steenberg}}{1982}]{Shull1982}
{Shull} J.~M.,  {van Steenberg} M.,  1982, \mn@doi [\apjs] {10.1086/190769},
  \href {http://adsabs.harvard.edu/abs/1982ApJS...48...95S} {48, 95}

\bibitem[\protect\citeauthoryear{{Smith} et~al.,}{{Smith}
  et~al.}{2017}]{grackle}
{Smith} B.~D.,  et~al., 2017, \mn@doi [\mnras] {10.1093/mnras/stw3291}, \href
  {http://adsabs.harvard.edu/abs/2017MNRAS.466.2217S} {466, 2217}

\bibitem[\protect\citeauthoryear{{Stanway}, {Eldridge}  \& {Becker}}{{Stanway}
  et~al.}{2016}]{Stanway2016}
{Stanway} E.~R.,  {Eldridge} J.~J.,   {Becker} G.~D.,  2016, \mn@doi [\mnras]
  {10.1093/mnras/stv2661}, \href
  {http://adsabs.harvard.edu/abs/2016MNRAS.456..485S} {456, 485}

\bibitem[\protect\citeauthoryear{{Sutherland} \& {Dopita}}{{Sutherland} \&
  {Dopita}}{1993}]{Sutherland1993}
{Sutherland} R.~S.,  {Dopita} M.~A.,  1993, \mn@doi [\apjs] {10.1086/191823},
  \href {http://adsabs.harvard.edu/abs/1993ApJS...88..253S} {88, 253}

\bibitem[\protect\citeauthoryear{{Teyssier}}{{Teyssier}}{2002}]{Teyssier2002}
{Teyssier} R.,  2002, \mn@doi [\aap] {10.1051/0004-6361:20011817}, \href
  {http://adsabs.harvard.edu/abs/2002A%26A...385..337T} {385, 337}

\bibitem[\protect\citeauthoryear{{Umeda} \& {Nomoto}}{{Umeda} \&
  {Nomoto}}{2002}]{Umeda2002}
{Umeda} H.,  {Nomoto} K.,  2002, \mn@doi [\apj] {10.1086/323946}, \href
  {http://adsabs.harvard.edu/abs/2002ApJ...565..385U} {565, 385}

\bibitem[\protect\citeauthoryear{{Vogelsberger}, {Genel}, {Sijacki}, {Torrey},
  {Springel}  \& {Hernquist}}{{Vogelsberger} et~al.}{2013}]{Volgelsberger2013}
{Vogelsberger} M.,  {Genel} S.,  {Sijacki} D.,  {Torrey} P.,  {Springel} V.,
  {Hernquist} L.,  2013, \mn@doi [\mnras] {10.1093/mnras/stt1789}, \href
  {http://adsabs.harvard.edu/abs/2013MNRAS.436.3031V} {436, 3031}

\bibitem[\protect\citeauthoryear{{Weingartner} \& {Draine}}{{Weingartner} \&
  {Draine}}{2001}]{Weingartner2001}
{Weingartner} J.~C.,  {Draine} B.~T.,  2001, \mn@doi [\apjs] {10.1086/320852},
  \href {http://adsabs.harvard.edu/abs/2001ApJS..134..263W} {134, 263}

\bibitem[\protect\citeauthoryear{{Wentzel}}{{Wentzel}}{1971}]{Wentzel1971}
{Wentzel} D.~G.,  1971, \mn@doi [\apj] {10.1086/150794}, \href
  {http://adsabs.harvard.edu/abs/1971ApJ...163..503W} {163, 503}

\bibitem[\protect\citeauthoryear{{Wiersma}, {Schaye}  \& {Smith}}{{Wiersma}
  et~al.}{2009a}]{Wiersma2009}
{Wiersma} R.~P.~C.,  {Schaye} J.,   {Smith} B.~D.,  2009a, \mn@doi [\mnras]
  {10.1111/j.1365-2966.2008.14191.x}, \href
  {http://adsabs.harvard.edu/abs/2009MNRAS.393...99W} {393, 99}

\bibitem[\protect\citeauthoryear{{Wiersma}, {Schaye}, {Theuns}, {Dalla Vecchia}
   \& {Tornatore}}{{Wiersma} et~al.}{2009b}]{Wiersma2009b}
{Wiersma} R.~P.~C.,  {Schaye} J.,  {Theuns} T.,  {Dalla Vecchia} C.,
  {Tornatore} L.,  2009b, \mn@doi [\mnras] {10.1111/j.1365-2966.2009.15331.x},
  \href {http://adsabs.harvard.edu/abs/2009MNRAS.399..574W} {399, 574}

\bibitem[\protect\citeauthoryear{{Wolfire}, {McKee}, {Hollenbach}  \&
  {Tielens}}{{Wolfire} et~al.}{2003}]{Wolfire2003}
{Wolfire} M.~G.,  {McKee} C.~F.,  {Hollenbach} D.,   {Tielens} A.~G.~G.~M.,
  2003, \mn@doi [\apj] {10.1086/368016}, \href
  {http://adsabs.harvard.edu/abs/2003ApJ...587..278W} {587, 278}

\bibitem[\protect\citeauthoryear{{Woosley} \& {Weaver}}{{Woosley} \&
  {Weaver}}{1995}]{Woosley1995}
{Woosley} S.~E.,  {Weaver} T.~A.,  1995, \mn@doi [\apjs] {10.1086/192237},
  \href {http://adsabs.harvard.edu/abs/1995ApJS..101..181W} {101, 181}

\makeatother
\end{thebibliography}


\bsp	
\label{lastpage}
\end{document}